\documentclass[a4paper,fleqn,usenatbib,useAMS]{mnras}
\pdfoutput=1
\usepackage{graphicx}
\usepackage{epstopdf}
\usepackage{subfig}
\usepackage{float}
\usepackage{xcolor}

\title[Inverted spectrum extragalactic radio sources]{GMRT observations of extragalactic radio sources with steeply inverted
spectra}

\author[Mukul-Mhaskey et al.]
  {Mukul Mhaskey$^{1}$\thanks{E-mail:mhaskeymukul@gmail.com},
   Gopal-Krishna$^{2}$,
   Pratik Dabhade$^{3,4}$,
   Surajit Paul$^{1}$, 
\newauthor
   Sameer Salunkhe$^{1}$ and
   S.K. Sirothia$^{5,6}$ 
\\
$^{1}$Department of Physics, Savitribai Phule Pune Unversity, Ganeshkhind, Pune 411007, India\\
$^{2}$Aryabhatta Research Institute of Observational Sciences (ARIES), Manora Peak, Nainital $-$ 263129, India\\
$^{3}$Inter University Centre for Astronomy and Astrophysics (IUCAA), Pune 411007, India\\
$^{4}$Leiden Observatory, Leiden University, Niels Bohrweg 2, 2333 CA, Leiden, Netherlands\\
$^{5}$Square Kilometre Array South Africa, 3rd Floor, The Park, Park Road, Pinelands, 7405, South Africa\\
$^{6}$Department of Physics and Electronics, Rhodes University, PO Box 94, Grahamstown, 6140, South Africa}
\pubyear{2018}

\begin{document}
\label{firstpage}
\pagerange{\pageref{firstpage}--\pageref{lastpage}}
\maketitle

\begin{abstract}
We report quasi-simultaneous GMRT observations of seven extragalactic radio sources at 150, 325, 610 and 1400 MHz, in an attempt to accurately define their radio continuum spectra, particularly at frequencies below the observed spectral turnover. We had previously identified these sources as candidates for a sharply inverted integrated radio spectrum whose slope is close to, or even exceeds $\alpha_c$ = +2.5, the theoretical limit due to synchrotron self-absorption (SSA) in a source of incoherent synchrotron radiation arising from relativistic particles with the canonical (i.e., power-law) energy distribution. We find that four out of the seven candidates have an inverted radio spectrum with a slope close to or exceeding +2.0, while the critical spectral slope $\alpha_c$ is exceeded in at least one case. These sources, together with another one or two reported in very recent literature, may well be the archetypes of an extremely rare class, from the standpoint of violation of the SSA limit in compact extragalactic radio sources. However, the alternative possibility that free-free absorption is responsible for their ultra-sharp spectral turnover cannot yet be discounted.  
\end{abstract}


\section{Introduction}
A few years ago we initiated a targeted search for \lq Extremely Inverted Spectrum Extragalactic Radio Sources\rq (EISERS) characterized by an integrated radio spectrum which turns over, attaining a slope $\alpha$ that exceeds $\alpha_{c}$ = +2.5 \citep{GopalKrishna2014}; (hereafter Paper I). This critical value, $\alpha_{c}$ is important as it represents the theoretical limit which can be achieved (via self-absorption) for an source of incoherent synchrotron radiation arising from a power-law energy distribution of 
relativistic charged particles, which is the basic mechanism widely held responsible for the radio continuum emission from active galactic nuclei \citep{Slish1963,Scheuer1968,Rybicki1979}. 
The Synchrotron Self-Absorption (SSA) can also produce shallower spectral gradient below the spectral turnover, as witnessed in many extragalactic radio sources, but this 
can be readily understood in terms of inhomogeneity within the source region contributing to the emission below the spectral turnover \citep{Odea1998,Tingay2003}. The basic physics underlying 
the SSA limit is that the source cannot have a brightness temperature in excess of the plasma temperature of the incoherently radiating non-thermal electrons \citep{Kellermann1969, Pacholczyk1970}. Discovery of rare radio galaxies with $\alpha$ $>$ $+$2.5 may then call for an explanation other than the SSA. One possibility 
is that free-free absorption (FFA) due to clouds/screen of thermal plasma is causing the sharp inversion of the integrated radio spectrum \citep{Kellermann1966,Kuncic1998,Kameno2000,
Vermeulen2003}. Different versions of this basic scenario have been summarized by \citet{Callingham2015}. Thus, FFA could occur in gas clouds of the narrow-line regions (NLR) engulfed by the 
radio lobes. AGN radiation and the synchrotron UV emission from the lobes could keep the compressed surface layers of the NLR clouds photo-ionized \citep{Stawarz2008,Ostorero2010}. The thermal gas 
clouds could also be photo-ionised by the bow shock associated with the jet propagating in the external medium \citep{vanBreugel1984, Bicknell1997, Bicknell2018}. As mentioned in Paper I, FFA effects
(leading to $\alpha$ $>$ $+$2.5) have indeed been 
observed in a few radio galaxies, albeit only for their parsec-scale nuclear radio jets. Prominent examples include the well-known radio source 3C 345 \citep{Matveenko1990}, Centaurus A 
\citep{Jones1996,Tingay2001}, Cygnus A \citep{Krichbaum1998}, NGC 1275/Perseus A \citep{Levinson1995,Walker2000}, NGC 4261 \citep{Jones2001} and NGC 1052 \citep{Vermeulen2003,Kadler2004}. 

An alternative scenario for the putative EISERS ($\alpha$ $>$ $+$2.5), perhaps more salient from the perspective of AGN physics, would be that in some rare sources, the low-energy spectrum of the 
radiating leptons itself differs from the canonical power-law shape, for example, being either mono-energetic, or a Maxwellian \citep{Rees1967}, or it has a large leptonic excess at low energies, over a 
power-law distribution
\citep{deKool1989}. On the observational side, only a few reports had existed of radio galaxies with integrated spectrum exhibiting a slope which approaches even $+$2.0. For instance, during a huge 
radio flare, the blazar-like spiral galaxy III Zw 2 exhibited an inverted spectrum with $\alpha$ $=$ +1.9$\pm$0.1 \citep{Falcke1999}. Another few possible examples showing sharply 
inverted integrated radio spectra are reported by \citet{Murphy2010} see, also \citep{Dallacasa2000}. But, since the SSA limit was not violated in any of these cases, this motivated us to initiate 
a systematic search for more extreme cases of inverted radio spectrum (Paper I).

It may be noted that, as compared to the opaque part of the radio spectrum, 
the origin of ultra-steep spectrum in the optically thin spectral regime has been 
discussed in the literature much more extensively. Such sources (having $\alpha < -1.0$) are 
commonly referred to as \lq\lq Ultra-Steep Spectrum \rq\rq~(USS) radio sources. With 
the increased availability of measurements below 0.5 GHz, several of 
them have, in fact, been reclassified as Gigahertz-Peaked-Spectrum (GPS)
sources \citep[e.g.][and references therein]{Callingham2017}, or even
Megahertz-Peaked-Spectrum (MPS) sources \citep{Falcke2004, Coppejans2015, Coppejans2016}. 
It has been suggested \citep{Callingham2017, Murgia2002} that in the case of USS
sources, the low-frequency part of the radio spectrum, which
is supposed to manifest the particle injection spectrum, remains 
essentially unobserved since it has already migrated either to 
frequencies that lie in the optically thick spectral regime, or even outside the 
normally accessed radio window. In this scenario, the observed ultra-steep 
spectrum in the optically thin regime is actually the radiatively 
steepened part of the radio spectrum, as expected from ageing 
of relativistic electrons in the continuous injection model 
\citep{Kardashev1962, Kellermann1964}. This interpretation of the 
USS spectrum being an ageing effect would also be consistent
with the empirically established {\it anti-correlation}, 
between the radio spectral curvature and the
steepness of the spectral slope measured,  
for classical double radio sources (say near 1 GHz in the rest-frame) \citep{Mangalam1995}. 
The same anti-correlation is, in fact, 
reflected in the observed straightness of the radio spectra of the USS sources, 
as highlighted by \citet{Klamer2006}. An alternative scenario to this 
which associates ultra-steep spectrum with aged but still active radio sources 
\citep[e.g.][]{Callingham2017, Mangalam1995}, posits that the USS sources are 
young and reside in dense environments (which is expected to be more common 
at high redshifts). As a result, the advance of their hot spots is 
slowed down, giving rise to a steep energy spectrum of the 
relativistic electrons accelerated there via the first-order Fermi
process, which causes an ultra-steep straight synchrotron
radio spectrum {\it ab initio} \citep{Athreya1998, Klamer2006, Bicknell1997}. 
This potentially attractive theoretical scenario would, however, need further substantiation, 
particularly in view of the empirical study presented by \citet{GopalKrishna2012}, 
which found no correlation of the particle injection spectral 
index with rotation measure for powerful double radio sources.

In section 2 we describe the selection procedure for the seven EISERS candidates. Section 3 contains the details of the quasi-simultaneous radio observations and the data analysis procedure. 
Notes on individual sources are given in section 4. This is followed by a brief discussion on the EISERS and the conclusions, in sections 5 \& 6 respectively.  

\section{Sample Definition}
As a first step, we reported in Paper I a set of 7 sources for which we had estimated an $\alpha > +$2.0, by comparing the TIFR.GMRT.SKY.Survey (TGSS/DR5) made at 150 MHz using the Giant Metrewave 
Radio Telescope (GMRT) \citep{Swarup1991} and the \lq Westerbork In the Southern Hemisphere \rq (WISH) survey at 352 MHz \citep{DeBreuck2002}. At that time these two surveys were the deepest available 
large-area sky surveys at metre wavelengths, with arc-minute or better resolution and a typical rms noise under 10 mJy$/$beam. Focussing attention only on the 7056 sources within the region of 
overlap between these two surveys, which had been listed as type \lq S \rq (i.e. single component) sources with flux densities above 100 mJy at 352 MHz in the  WISH catalogue, we found seven sources 
having $\alpha$ (150-352 MHz) $> +$2.0 (Paper I). Two of them were actually undetected at 150 MHz and hence only lower limits could be set to the slopes of their inverted spectra. All the seven sources 
appear unresolved in their TGSS images and a GPS type peaked radio spectrum could be confirmed for at least 5 of the 7 sources, based on the existing radio measurements.\footnote{GPS sources are 
defined as having an integrated spectrum that shows a single peak and steep slopes on either side of the peak, cf. \citep{Spoelstra1985, GopalKrishna1983, GopalKrishna1993, Odea1998, An2012}}. 

In Paper I we underlined two of the seven sources as highly promising EISERS candidates, since their non-detection in the TGSS at 150 MHz implied $\alpha$ (150-352 MHz) $> +$2.35. Soon thereafter, 
\citet{Callingham2015} reported discovery of a very interesting \lq extreme GPS source\rq~PKS B0008$-$421, based on an exceptionally dense spectral sampling between 0.1 and 22 GHz, covering 
comprehensively both sides of the single spectral peak seen near 0.6 GHz. Strikingly, its spectral slope is seen to become as large as $+$2.4 below the turnover frequency, closely approaching the 
SSA limit of $\alpha_c = +2.5$. Although, the multi-frequency measurements used by them are non-contemporaneous, the authors argued that any flux variability should be too small to have 
significantly affected the radio spectrum below the turnover. Based on physical considerations, they showed a slight preference for an inhomogeneous FFA model, over the SSA model, as being the 
dominant cause of the sharp spectral turnover observed. In a follow-up paper, \citet{Callingham2017} (hereafter CEG17) have reported a few more examples of such sources, which are briefly discussed 
in Sect. 5.

Clearly, the results presented in Paper I were only meant to be an initial step towards a systematic search for EISERS ($\alpha > +$2.5), particularly in view of the following two caveats mentioned there:
Firstly, the two radio surveys (TGSS/DR5 and WISH) used for computing the spectral indices of individual sources were made at epochs nearly a decade apart. The long time interval could then 
have introduced significant uncertainty due to flux variability expected from refractive interstellar scintillation at such low frequencies \citep{Bell2019}.
The second caveat is that since the overlap region between the two surveys lies at fairly low declinations ($-$10 $>$ $\delta$ $>$ $-$30 deg), the north-south beamwidth of the 352 MHz WISH survey 
is rather large (FWHM $=$ 0.9 cosec $\delta$ $\sim$ 2 arcmin). The enhanced confusion could then have significantly impacted the accuracy of flux measurement. It may also be noted that even the 
WENSS \citep{Rengelink1997}, which is the northern-sky counterpart of WISH and therefore much less affected by the beam elongation, is known to be off the \citet{Baars1977} flux scale or in the flux scale of Roger, Costain \& Bridle, \citep[RCB,][]{Roger1973} by over 10\% \citep[see][]{Hardcastle2016}. Therefore, the combined uncertainty associated with the measurement and calibration of the WISH and TGSS-DR5 flux densities could have biased the previous estimates 
of spectral index. This also explains why the WISH flux densities are often found to be significantly off from the GMRT measurements at 325 MHz (Section 4). Rectifying both these 
shortcomings, we present here previously unpublished GMRT observations of all the seven EISERS candidates reported in Paper I. These observations, made at 150, 325, 610 and 1400 MHz, have the highest 
sensitivity and resolution currently available for these sources at such low frequencies and, furthermore, they are nearly contemporaneous, all having been carried out within a time span of just four 
weeks (3 weeks between the observations at the lowest two frequencies, namely 150 and 325 MHz). 
We further note that the present observations have provided for each source at least two well-spaced data points below the 
spectral peak, thus enabling a better constrained spectral turnover than is the norm for GPS sources, a point also highlighted in CEG17.    

\section{Radio Observation and Analysis}
\subsection{Radio Observations}
The 7 EISERS candidates were observed  with the GMRT \citep{Swarup1991} in the snapshot mode, quasi-simultaneously at 150, 325, 610 and 1400 MHz, between 2014 July 09 and 2014 August 05 
(Table~\ref{table:obs-log}). The integration time was 2 sec at 150 MHz (bandwidth 16 MHz) and 16 sec at the remaining 3 frequencies (bandwidth 32 MHz). One flux calibrator, out of 3C 286, 3C 48, 
and 3C 147, was observed at the start and the end of each observing session. Their flux densities were taken from the VLA calibrator manual \citep{Perley2017}. Each snapshot of a given target 
source was sandwiched between a pair of snapshots on its phase calibrator(s). The average total on-target time depends on frequency, being $\sim$40 minutes at 150 MHz and as short as 2 minutes at 
1400 MHz. Table~\ref{table:obs-log} contains the observation log providing additional details for the different observing frequencies. Table~\ref{table:spec-prop} lists flux density measurements 
at different frequencies, taken from the present GMRT observations and from the literature. The radio contour maps at 150 MHz and 325 MHz for the sources J0242$-$1649 and J1209$-$2032 are presented in 
Figures~\ref{fig:J0242} \&~\ref{fig:J1209}, respectively.

\begin{table*}
\small
\caption{GMRT observation log and the map parameters.\\}
\label{table:obs-log}
\begin{tabular}{cccccccc}\\
\hline

\multicolumn{1}{c}{Source} & Obs. date & \multicolumn{1}{c}{No. of} & \multicolumn{1}{c}{Total duration} & 
\multicolumn{1}{c}{Phase} & \multicolumn{1}{c}{Synthesised beam (FWHM)} & \multicolumn{1}{c}{Position angle} & 
\multicolumn{1}{c}{rms noise} \\
\multicolumn{1}{c}{} &  & \multicolumn{1}{c}{snapshots} & \multicolumn{1}{c}{(minutes)} & \multicolumn{1}{c}{calibrators} 
& \multicolumn{1}{c}{(arc sec)} & \multicolumn{1}{c}{of the synthesised beam} & \multicolumn{1}{c}{(mJy/beam)} \\

\hline
\textbf{150 MHz}\\
J0242$-$1649 & 2014 Jul 09 & 4 & 50 & 0116$-$208  &23.7 $\times$ 16.7 &$-$33$^\circ$ & 1.3\\
J0442$-$1826 & 2014 Jul 09 & 3 & 60 & 0409$-$179  &27.2 $\times$ 17.4 &$-$34$^\circ$ & 1.2\\
J1003$-$2514 & 2014 Jul 11 & 2 & 40 & 0837$-$198  &24.9 $\times$ 16.4 &$-$04$^\circ$ & 3.4\\
J1031$-$2228 & 2014 Jul 09 & 2 & 40 & 1033$-$343  &24.6 $\times$ 15.7 &07$^\circ$ & 3.1\\
J1207$-$2446 & 2014 Jul 09 & 2 & 35 & 1311$-$222  &33.7 $\times$ 12.8 &42$^\circ$& 3.6\\
J1209$-$2032 & 2014 Jul 09 & 2 & 35 & 1311$-$222  &31.9 $\times$ 13.4 &47$^\circ$ & 6.5\\ 
J1626$-$1127 & 2014 Jul 09 & 2 & 40 & 1822$-$096  &27.3 $\times$ 15.1 &17$^\circ$ & 7.5\\
\\ 
\textbf{325 MHz}\\
J0242$-$1649 & 2014 Aug 01 & 4 & 35 & 0409$-$179  &14.1 $\times$ 07.0 & $-$20$^\circ$ &0.2 \\
J0442$-$1826 & 2014 Aug 01 & 3 & 25 & 0409$-$179  &13.7 $\times$ 06.9 & 02$^\circ$ &0.3 \\
J1003$-$2514 & 2014 Aug 01 & 2 & 10 & 0837$-$198  &24.6 $\times$ 07.6 & $-$48$^\circ$ &0.5 \\
J1031$-$2228 & 2014 Aug 01 & 1 & 10 & 0837$-$198  &23.1 $\times$ 08.2 & $-$51$^\circ$ &0.7 \\
J1207$-$2446 & 2014 Aug 01 & 1 & 05 & 1311$-$222  &18.1 $\times$ 07.3 & 50$^\circ$ &0.5 \\
J1209$-$2032 & 2014 Aug 01 & 1 & 05 & 1311$-$222  &17.7 $\times$ 07.1 & 54$^\circ$ &0.7 \\
J1626$-$1127 & 2014 Aug 01 & 2 & 10 & 1822$-$096  &11.7 $\times$ 06.8 & 38$^\circ$ &0.6 \\
\\
\textbf{610 MHz}\\
J0242$-$1649 & 2014 Aug 05 & 2 & 10 & 0409$-$179 &07.18 $\times$ 04.92 & 55$^\circ$ &0.4 \\
J0442$-$1826 & 2014 Aug 05 & 2 & 10 & 0409$-$179 &05.72 $\times$ 05.57 & 74$^\circ$ &0.5 \\
J1003$-$2514 & 2014 Jul 18 & 2 & 4  & 0837$-$198 &08.39 $\times$ 03.56 & 36$^\circ$ &0.3 \\
J1031$-$2228 & 2014 Jul 18 & 2 & 4  & 0837$-$198 &07.36 $\times$ 03.61 & 33$^\circ$ &0.6 \\
J1207$-$2446 & 2014 Jul 18 & 2 & 4  & 1311$-$222 &07.20 $\times$ 04.14 & 30$^\circ$ &0.5 \\
J1209$-$2032 & 2014 Jul 18 & 2 & 5  & 1311$-$222 &06.69 $\times$ 04.22 & 35$^\circ$ &0.5 \\
J1626$-$1127 & 2014 Jul 18 & 2 & 5  & 1419$+$064 &07.16 $\times$ 04.86 & $-$72$^\circ$ &0.5 \\
\\
\textbf{1400 MHz}\\
J0242$-$1649 & 2014 Jul 23 & 1 & 2 & 0409$-$179 &03.94 $\times$ 02.02 &51$^\circ$ &0.2 \\
J0442$-$1826 & 2014 Jul 23 & 1 & 2 & 0409$-$179 &02.78 $\times$ 02.17 &10$^\circ$ &0.3 \\
J1003$-$2514 & 2014 Jul 23 & 1 & 5 & 0837$-$198 &05.22 $\times$ 01.97 & $-$48$^\circ$ &0.3 \\
J1031$-$2228 & 2014 Jul 23 & 1 & 5 & 0837$-$198 &04.79 $\times$ 01.91 & $-$51$^\circ$ &0.5 \\
J1207$-$2446 & 2014 Jul 22 & 1 & 2 & 1311$-$222 &11.01 $\times$ 02.65 & $-$02$^\circ$ &0.4 \\
J1209$-$2032 & 2014 Jul 22 & 1 & 2 & 1311$-$222 &09.97 $\times$ 02.73 & $-$07$^\circ$ &0.4 \\
J1626$-$1127 & 2014 Jul 22 & 1 & 2 & 1419$+$064 &08.76 $\times$ 03.14& $-$24$^\circ$ &0.3 \\
\\
\hline
\end{tabular}
\end{table*}

\subsection{Analysis}
The astronomical packages \textsc{spam} and \textsc{aips} were used for the data analysis. The visibilities at the low frequencies of 150 MHz and 325 MHz were processed using the \textsc{spam} (Source 
Peeling and Atmospheric Modelling; \citet{Intema2014}) package, which is a semi-automated pipeline based on \textsc{aips}, Parseltongue and Python. \textsc{spam} performs a series of iterative flagging 
and calibration and the imaging is done with direction dependent calibration. This package has recently been used by \citet{Intema2017} for processing the entire TIFR-GMRT SKY SURVEY (TGSS) data at 
150 MHz. Details of \textsc{spam} and its routines are provided by \citet{Intema2017}.

The 610 MHz and 1400 MHz GMRT observations were reduced using the Astronomical Image Processing System (\textsc{aips}). The standard procedure consisted of first cleaning up the data for RFI (bad 
visibility data points) using the \lq flgit\rq~and \lq uvflg\rq~tasks, followed by the flux and phase calibration protocols. The Perley-Butler \citep{Perley2017} absolute flux density scale was used 
to set the flux densities of the flux calibrators, which in turn were used to define the flux scale for the respective phase calibrators and the target sources. Calibrated visibilities were transformed 
into radio image (deconvolved images) using the \lq IMAGR\rq~task which performs the self-calibration for mitigating the antenna-based phase and amplitude errors. Several rounds of phase-only 
self-calibration cycles were carried out, concluded by one round of amplitude plus phase calibration. Lastly, a correction for the antenna primary beam was applied to the map obtained after the final 
self-calibration round. The flux density uncertainties of individual sources have been computed using the expression given in equation~\ref{eq:error}, wherein the 10\% of the peak flux density accounts 
for the systematic error component, including the small errors arising from the gain dependence of the antennas on their elevation angle changes \citep{Chandra2004}.

\begin{equation} \label{eq:error}
\sqrt{ \rm{(map~rms)}^{2} + (10 \%~\rm{of~the~peak~flux})^{2}}
\end{equation}

\begin{table*}
\tiny\addtolength{\tabcolsep}{-4pt}
\caption{Positions (J2000), flux densities and spectral indices of the 7 EISERS candidates.}
\label{table:spec-prop}
\begin{tabular}{cccccccccccccc}\\
\hline

\multicolumn{1}{c}{Source position} & \multicolumn{1}{c}{150 MHz} & \multicolumn{1}{c}{325 MHz} & \multicolumn{1}{c}{610 MHz} & \multicolumn{1}{c}{1400 MHz} & \multicolumn{1}{c}{150 MHz}& \multicolumn{1}{c}{151 MHz}& \multicolumn{1}{c}{352 MHz}& \multicolumn{1}{c}{1400 MHz} & \multicolumn{1}{c}{4.85 GHz}& \multicolumn{1}{c}{4.85 GHz} &\multicolumn{1}{c}{8 GHz} & \multicolumn{1}{c}{20 GHz} & \multicolumn{1}{c}{Spectral Index}\\
GMRT$^{\star}$ & GMRT & GMRT & GMRT & GMRT & TGSS-ADR1$^{\star}$ & GLEAM$^{\star}$ & WISH$^{\star}$ &NVSS$^{\star}$ & PMN$^{\star}$ &ATCA$^{\star}$ &ATCA$^{\star}$ & ATCA$^{\star}$ & (150-352 MHz)\\
 at 325 MHz&(mJy)&(mJy)&(mJy)&(mJy)&(mJy)&(mJy)&(mJy)&(mJy)&(mJy)&(mJy)&(mJy)&(mJy)&\\

\hline
$ $02 42 10.57  & 6.4${\pm}$1.4 &62.7${\pm}$6.3 &167.2${\pm}$16.7 & 75.2${\pm}$7.5 & $<$13& $<$54.9 &106${\pm}$4.5 &96.7${\pm}$2.9 & & & & & 2.95${\pm}$0.31\\
$-$16 49 33.5   &(1.3) &(0.2) &(0.4) &(0.2) &(2.6) &(18.3) & & & & & & &\\
\\
$ $04 42 01.20  &35.6${\pm}$3.8 & 88.2${\pm}$8.8 &105.8${\pm}$10.6& 43.5${\pm}$4.5 &$<$16 &$<$73.2 &105${\pm}$4.4 &50.9${\pm}$1.6 &85${\pm}$11 & & & &1.17${\pm}$0.19 \\
$-$18 26 34.0   &(1.3) &(0.3) &(0.5) &(0.3) &(3.2) &(24.4) & & & & & & &\\
\\
$ $10 03 06.19  & 9.4${\pm}$3.5& 47.9${\pm}$4.8 &154.8${\pm}$15.5& 62.1${\pm}$6.2 &19.2${\pm}$2.0 &37.5${\pm}$20.6 &143${\pm}$6.1 &74.1${\pm}$2.8 & & & & &2.11${\pm}$0.50 \\
$-$25 14 05.5   &(3.4) &(0.5) &(0.3) &(0.3) &(4.1) &(20.8) & & & & & & &\\
\\
$ $10 31 52.20  & 22.8${\pm}$3.6& 107.1${\pm}$10.7 &311.6${\pm}$31.2& 307.7${\pm}$30.8& 56.1${\pm}$9.5 &46.0${\pm}$17.1 &191${\pm}$7.1 &371.4${\pm}$11.2 &328${\pm}$20 &371${\pm}$19& 291${\pm}$15&124${\pm}$8 &2.00${\pm}$0.24\\
$-$22 28 26.2   &(3.1) &(0.7) &(0.6) &(0.5) &(4.7) &(20.3) & & & & & & & \\
\\
$ $12 07 05.29  & & 172.3${\pm}$17.2 &424.4${\pm}$42.44& 205.2${\pm}$20.52 &69.6${\pm}$9.2 &73.4${\pm}$19.3 &380${\pm}$23 &226.7${\pm}$6.8 &105${\pm}$12 & & & &1.17${\pm}$0.21$^{\ddagger}$ \\
$-$24 46 28.7   & &(0.5) &(0.5) &(0.4) &(3.6) &(22.6) & & & & & & &\\
\\
$ $12 09 15.31$^{\dagger}$  &$<$19.5 & 149.4${\pm}$14.9& 422.9${\pm}$42.3& 326.2${\pm}$32.6 &$<$17.5 &$<$64.5 &207${\pm}$8.4 &353.7${\pm}$10.6 &573${\pm}$32&832${\pm}$42 &1177${\pm}$61 &707${\pm}$46 &$>$2.64${\pm}$0.13\\
$-$20 32 34.4  &(6.5) &(0.7) &(0.5) &(0.4) &(3.5) &(21.5) & & &  & & & &\\
\\
$ $16 26 51.82   & 37.8${\pm}$8.4&  152.5${\pm}$15.3& 154.4${\pm}$15.5& 56.7${\pm}$5.7 &84.1${\pm}$9.5 &65.1${\pm}$41.4 &206${\pm}$8.5 &52.3${\pm}$1.6 & & & & & 1.80${\pm}$0.32\\
$-$11 27 24.0   &(7.5) &(0.6) &(0.5) &(0.3) &(3.3) &(47.6) & & & & & &\\
\hline
\end{tabular}

{$^{\star}$ References: GMRT -- \citet{Swarup1991}; TGSS-ADR1 --\citet{Intema2017}; GLEAM -- \citet{Hurley-Walker2017}; 
WISH -- \citet{DeBreuck2002}; NED -- NASA Extragalactic Database; NVSS -- \citet{Condon1998}; PMN -- \citet{Griffith1994}; 
ATCA -- \citet{Murphy2010}.} \\
{Note that the values given inside parentheses, just below the flux densities, are the rms errors of the respective maps.} \\
{$^{\dagger}$ J1209$-$2032 is identified with a galaxy at $z$ = 0.404 \citep{Healey2008}}\\
{$^{\ddagger}$ Unlike the other values given in this column, calculation of this spectral index uses the 150 MHz flux 
density taken from TGSS-ADR1, since the present 150 MHz GMRT map is too noisy.}
\end{table*}

As a cautionary check on possible systematic bias in the flux density scales of the present 150 MHz GMRT maps, we have additionally determined the scaling factors by comparing the present map of
each target field with its counterparts in the TGSS-ADR1 \citep{Intema2017} and/or GLEAM surveys \citep{Hurley-Walker2017}, also made at 150 MHz. For a given GMRT map (measuring 1 $\times$ 1 deg), the 
flux scaling factor (FSF) was determined by taking average of the ratios of the measured flux densities of several discrete sources in the present 150 MHz GMRT map and their counterparts in the 
other 150 MHz map being compared. This was done after ensuring that the chosen sources are stronger than $\sim$100 mJy and located in relatively isolated environment within the maps. 
Table~\ref{table:FSF} gives the so obtained values of FSF for each of our 150 MHz GMRT map. Thus, multiplying the present 150 MHz GMRT flux densities with the FSF value(s) estimated for that field 
translates the present flux densities to the flux scales of the TGSS-ADR1 and the GLEAM survey. The correspondingly adjusted values of the spectral index $\alpha$ (150-325 MHz) of individual target 
sources are also listed in Table~\ref{table:FSF}. We caution that the estimates of FSF have only been used for calculating the spectral indices $\alpha_{(1)}$ \& $\alpha_{(2)}$ given in 
Table~\ref{table:FSF} and are not meant for general applicability. Figure~\ref{fig:spec_all} shows the radio spectra of the seven sources.     

\begin{table}
\tiny\addtolength{\tabcolsep}{-3pt}
\caption{Flux-density scaling factors (FSFs) and the correspondingly adjusted values of $\alpha$ (150$-$325 MHz).} 
\label{table:FSF}
\begin{tabular}{cccccc}\\
\hline
\multicolumn{1}{c}{Source} & \multicolumn{1}{c}{FSF-1} & \multicolumn{1}{c}{FSF-2} & \multicolumn{1}{c}{$\alpha^{\dagger}$} 
& \multicolumn{1}{c}{$\alpha_{TGSS}^{\ddagger}$} & \multicolumn{1}{c}{$\alpha_{GLEAM}^{\ddagger}$} \\
\multicolumn{1}{c}{} & \multicolumn{1}{c}{TGSS/GMRT} & \multicolumn{1}{c}{GLEAM/GMRT} & \multicolumn{1}{c}{} & 
\multicolumn{1}{c}{} \\
\hline
$ $02 42 10.57  &0.95 &0.83& 2.95${\pm}$0.31&3.02${\pm}$0.33 &3.19${\pm}$0.37 \\
$-$16 49 33.5   & & & & \\
\\
$ $04 42 01.20  &0.73 &0.72& 1.17${\pm}$0.19&1.58${\pm}$0.19 &1.60${\pm}$0.19 \\
$-$18 26 34.0   & & & & \\
\\
$ $10 03 06.19  &1.06 &0.92& 2.11${\pm}$0.50&2.03${\pm}$0.48 &2.21${\pm}$0.54 \\
$-$25 14 05.5   & & & & \\
\\
$ $10 31 52.20  &1.21 &0.98& 2.00${\pm}$0.24&1.75${\pm}$0.24 &2.03${\pm}$0.26 \\
$-$22 28 26.2   & & & & \\
\\
$ $12 07 05.29  &$-$ &$-$& 1.17${\pm}$0.21$^{\star}$ &$-$ &$-$ \\
$-$24 46 28.7 	& & & & \\
\\
$ $12 09 15.31  &1.44 &1.37& 2.64${\pm}$0.13&2.16${\pm}$0.39 &2.23${\pm}$0.41 \\
$-$20 32 34.4   & & & & \\
\\
$ $16 26 51.82  &1.26 &0.97& 1.80${\pm}$0.32&1.51${\pm}$0.27 &1.84${\pm}$0.32 \\
$-$11 27 24.0   & & & & \\
\hline
\end{tabular}

{$^{\dagger}$ The present estimate of spectral Index (150-325 MHz) when no scaling factor is applied.
Flux densities at both 150 MHz and 325 MHz are measured from the GMRT maps reported here, except for J0242$-$1649 (see the footnote to Table 2).} \\
{$^{\star}$ The 150 MHz flux density for this source has been taken from TGSS-ADR1, as the present GMRT map is severely affected 
by RFI.} \\
{$^{\ddagger}$ Estimate of spectral Index (150-325 MHz) when scaling factors from TGSS and GLEAM, respectively are applied.} 
\end{table}

\section{Notes on individual sources}

\subsection*{J0242$-$1649}
\begin{figure}
\centering	
\includegraphics[width=\columnwidth]{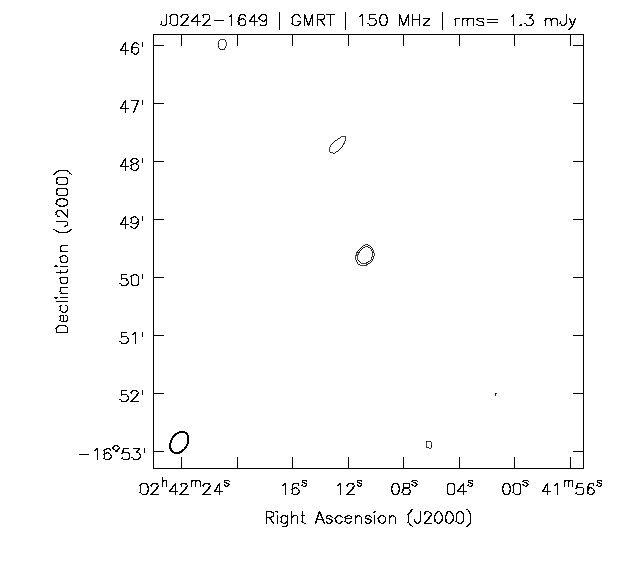} \\
\includegraphics[width=\columnwidth]{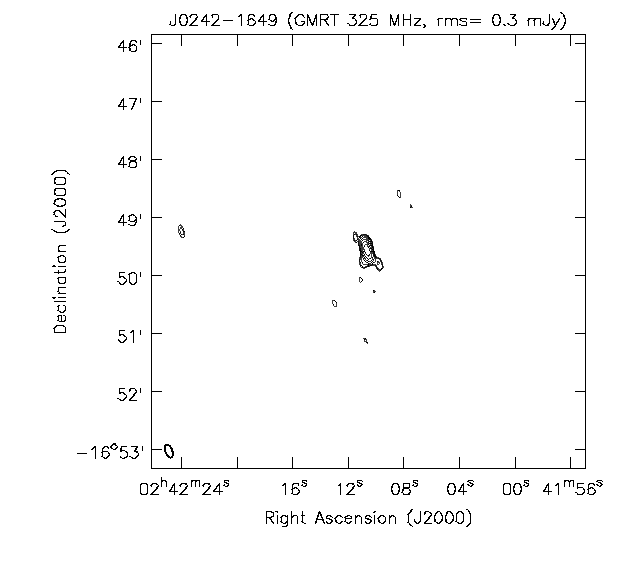}
\caption{GMRT contour maps of J0242$-$1649 at 150 MHz and 325 MHz, respectively. The contours are drawn at 3,4,8,16,32,64 \& 128 times the image rms noise which is 1.3 mJy at 150 MHz 
and 0.3 mJy at 325 MHz. The beam sizes are 23.7 $\times$ 16.7$"$ (PA= -33$\degr$) and 14.1 $\times$ 7.0$"$ (PA= $-$20$\degr$) at 150 MHz and 325 MHz, respectively. The target source lies at the centre 
of the map.} \label{fig:J0242}  
\end{figure}

Although this source is not seen in the TGSS-ADR1 and GLEAM maps, the high sensitivity of the present GMRT map ($\sigma$ $=$ 1.3 mJy, Figure~\ref{fig:J0242}) has led to its detection at 150 MHz. Its 
flux densities at 150 MHz and the 3 higher frequencies, measured in the present GMRT observations are given in Table~\ref{table:spec-prop}, along with the same information for the remaining six of our 
target sources. The new GMRT maps at 150 MHz and 325 MHz for our most promising targets are shown in Figures~\ref{fig:J0242} and ~\ref{fig:J1209} (the maps for the remaining sources are included as 
part of the online material). At 325 MHz, although the source is clearly detected in the present map, its flux density is much lower ($\sim$60\%) than that given in the WISH catalog at 352 MHz. Still, 
its inverted spectrum with $\alpha$$ = $+2.95${\pm}$0.31 remains steep enough to qualify it as an EISERS (Table~\ref{table:spec-prop}). 

\subsection*{J0442$-$1826}
The source is detected in the present 150 MHz GMRT map ($\sigma$ $\sim$ 1.2 mJy, Figure 2-online material), precisely at its NVSS and GMRT 325 MHz positions (Table~\ref{table:spec-prop}). However, the present flux density estimate at 150 MHz (35.6 mJy) is approximately twice that reported in the TGSS$/$DR5 (see Paper I). Curiously, no source is detected at this position in the TGSS-ADR1 map at 150 MHz ($\sigma \sim$3.2 mJy, Figure 1-online material). This is intriguing, particularly as all other sources of similar (and, of course higher) intensities have been detected in the two 150 MHz maps just mentioned, as is indeed expected. Unfortunately, the source is too weak to appear in the GLEAM map at 151 MHz ($\sigma$ $\sim$ 24 mJy). At 325 MHz, the present estimate of flux density (Table~\ref{table:spec-prop}) is only marginally lower than the WISH value at 352 MHz used in Paper I. The VLBI images at 4.3 and 8.4 GHz reveal a resolved structure consisting of a dominant flat-spectrum and a linear extension of size $\sim$ 20 mas towards south-east direction\footnote{http://astrogeo.org/vlbi$\_$images/}. 
 
\subsection*{J1003$-$2514}
With a flux density of 9.4${\pm}$3.4 mJy this source is detected at just 2.8 $\sigma$ level in the present 150 MHz GMRT map (Figure 3-online material). In the 150 MHz TGSS-ADR1 map ($\sigma$ $\sim$ 4.1 mJy) it is registered at a stronger level of 19.2 mJy. Note that the flux scales of the two maps are found to be in good agreement (Table~\ref{table:FSF}). In any case, with a spectral slope of +2.11${\pm}$0.50 (Table~\ref{table:FSF}), this source remains an interesting candidate to follow-up. We also note that its flux density of 143${\pm}$6.1 mJy in the WISH catalogue at 352 MHz, which was used in Paper I, is nearly a factor of 3 higher than the present estimate of 47.9${\pm}$4.8 mJy at 325 MHz. The large discrepancy could partly be due to uncertainty in the gain and absolute flux density calibrations, which is already over 10\% for the WENSS \citep{Hardcastle2016} and probably much more for WISH owing to its highly elongated beam. Since its nearest feature seen in the present 325 MHz GMRT map, which is offset towards north-west by $\sim$1.5 arcmin and is $\sim$ 4 times weaker than the target source (Figure 3-online material), any confusion due to it is unlikely to account for the huge excess of the WISH flux density.  

\subsection*{J1031$-$2228}
The source is clearly seen in the present 150 MHz GMRT map ($\sigma \sim$ 3.1 mJy, Figure 4-online material). However, its flux density (22.8${\pm}$3.6 mJy) is much below the TGSS-ADR1 value of 56.1${\pm}$9.5 at 150 MHz (Table~\ref{table:spec-prop}). We note that the flux scale of our 150 MHz map is consistent with the GLEAM map to within $\sim$ 2\%, but lower by $\sim$ 25\% in comparison to the 150 MHz TGSS-ADR1 map. However, even this falls too short of explaining the discrepancy. The large downward revision of flux density at 150 MHz is mirrored when the present value of $\sim$ 107.1${\pm}$10.7 mJy at 325 MHz is compared with the WISH estimate of 191${\pm}$7.1 mJy at 325 MHz which we used in Paper I. A similar, albeit less pronounced, trend is seen at 1.4 GHz, when the present estimate of 307.7 mJy is compared with the NVSS value of 371.4 mJy.
\citet{Taylor2009} have inferred a rotation measure of 52.6$\pm$18 rad m$^{-2}$ for this source, suggesting a moderate presence of magneto-ionic plasma. The AT20G catalogue reports fractional polarisation of 4.8\% at 20 GHz, 2.1\% at 8 GHz and 1.6\% at 5 GHz (\citet{Murphy2010}). The VLBA maps at 2.3, 4.3 and 7.6 GHz exhibit a roughly linear structure of overall size $\sim$ 20 mas, including a flat-spectrum central core and two lobes each resolved into peaks, the western lobe being the brighter one\footnote{http://astrogeo.org/vlbi$\_$images/}.

\subsection*{J1207$-$2446}
Unfortunately, its present GMRT 150 MHz observations were severely affected by the RFI and therefore no useful map could be obtained. Nonetheless, an adequately sensitive TGSS-ADR1 map is available at this frequency ($\sigma$ $\sim$ 3.6 mJy, Figure 5-online material), clearly showing the target source. A comparison with the GLEAM 150 MHz map shows that the TGSS-ADR1 flux densities are systematically low by a factor FSF $\sim 1.6$ (Table~\ref{table:FSF}). However, even if this is ignored, the source's spectral index is only ($\alpha$ $=$ 1.17${\pm}$0.21, i.e., too small to qualify it as an EISERS (Table~\ref{table:spec-prop}). Its present estimate at 325 MHz is $\sim$ 172.3${\pm}$17.2 mJy which is only about half the WISH value we had used in Paper I. This has led to a major downward revision of its spectral index and the source is no longer an EISERS candidate. 

\subsection*{J1209$-$2032}
\begin{figure}
\centering	
\includegraphics[width=\columnwidth]{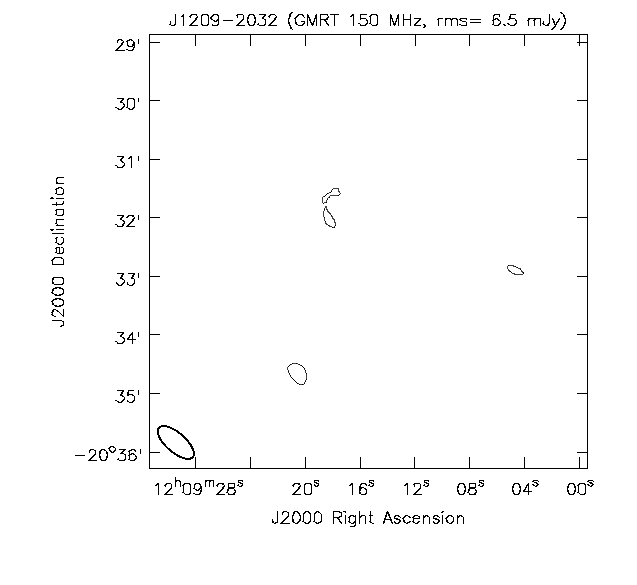} \\
\includegraphics[width=\columnwidth]{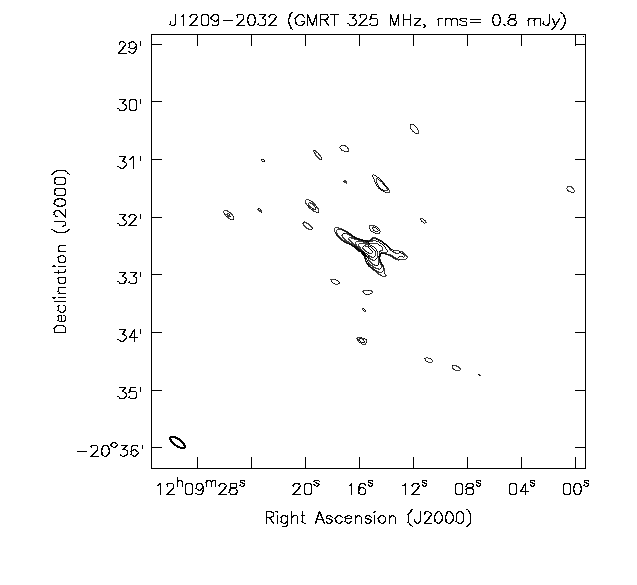} 
\caption{GMRT contour maps of J1209$-$2032 at 150 MHz and 325 MHz, respectively. The contours are drawn at 3,4,8,16,32,64 and 128 times the image rms noise which is 6.5 mJy at 150 MHz and 0.8 mJy at 325 MHz. The beam sizes are 31.9 $\times$ 13.4$"$ (PA= 47$\degr$) and 17.7 $\times$ 7.1$"$ (PA= 54$\degr$) at 150 MHz and 325 MHz, respectively. The target source lies at the centre of the map. The complex structure seen at 325 MHz map probably suffers from severe artefacts, arising from the exceptionally short duration (5 mins) for which this source could be observed (Table 1).}  
\label{fig:J1209} 
\end{figure}

The present 150 MHz GMRT map is comparatively noisy ($\sigma \sim 6.5$ mJy, Figure~\ref{fig:J1209}) near the field centre, yielding an upper limit of 19.5 mJy for this undetected source. A further revision of this limiting value to $\sim$ 27 mJy may be necessary since we find for this GMRT map a scaling factor of $\sim$ 1.4, relative to both TGSS-ADR1 and GLEAM maps (Table~\ref{table:FSF}). While this upper limit at 150 MHz is fully consistent with that given in Paper I, a large difference exists at 325 MHz, in the sense that the present GMRT estimate of 149${\pm}$14.9 mJy is significantly below the WISH catalog value of 207${\pm}$8.4 mJy at 352 MHz which we used in Paper I. The cause of this discrepancy is likely to be the same as mentioned above for the case of J1003$-$2514; it could be related to the presence of a few strong sources within the field (Figure~\ref{fig:J1209}) and the rather large, elongated beam of the WISH map (sect. 1). Using just the present GMRT flux densities at 150 MHz and 325 MHz, gives a spectral index of $>$+2.64${\pm}$0.13, making this source a possible case of EISERS. The other two estimates, $\alpha_{(1)}$ \& $\alpha_{(2)}$ (Table 3) are somewhat smaller, and hence could well be consistent with the SSA limit. \citet{Healey2008} have identified this source with a galaxy at $z$ = 0.404. The AT20G catalogue reports fractional polarisation of 1\% at 20 GHz, 8 GHz and 5 GHz (\citet{Murphy2010}). The VLBI images at 2.3, 4.3 and 7.6 GHz reveal a pair of lobes separated by $\sim$ 25 mas, and straddling a flat-spectrum core \footnote{http://astrogeo.org/vlbi$\_$images/}.

\subsection*{J1626$-$1127}
The source is clearly detected in the present 150 MHz GMRT map ($\sigma \sim 7.5$ mJy, Figure 6-online material), with a flux density of 37.8${\pm}$8.4 mJy. This is consistent with the TGSS/DR5 value 
used in Paper I. Unlike the present 150 MHz GMRT map, the source is seen to have some resolved emission in the TGSS-ADR1 map ($\sigma$ = 3.3 mJy ) and has a much higher integrated flux density 
($\sim$84.1${\pm}$9.5 mJy at 150 MHz). No other discrepancy is evident between the two maps, except for the flux scaling factor which is somewhat large (FSF $\sim$ 1.26, Table~\ref{table:spec-prop}). 
At 325 MHz, the source is seen at a significantly lower level in the present GMRT map (152.5${\pm}$15.3 mJy), as compared to the WISH value of 206${\pm}$8.5 mJy at 352 MHz which we used in Paper I. 
But in either case, its spectral index remains below $+$2.0 (Table~\ref{table:spec-prop}), ruling it out as an EISERS candidate.

\begin{figure*}
\centering
\hbox{\hspace{13.0em} \includegraphics[scale=0.65]{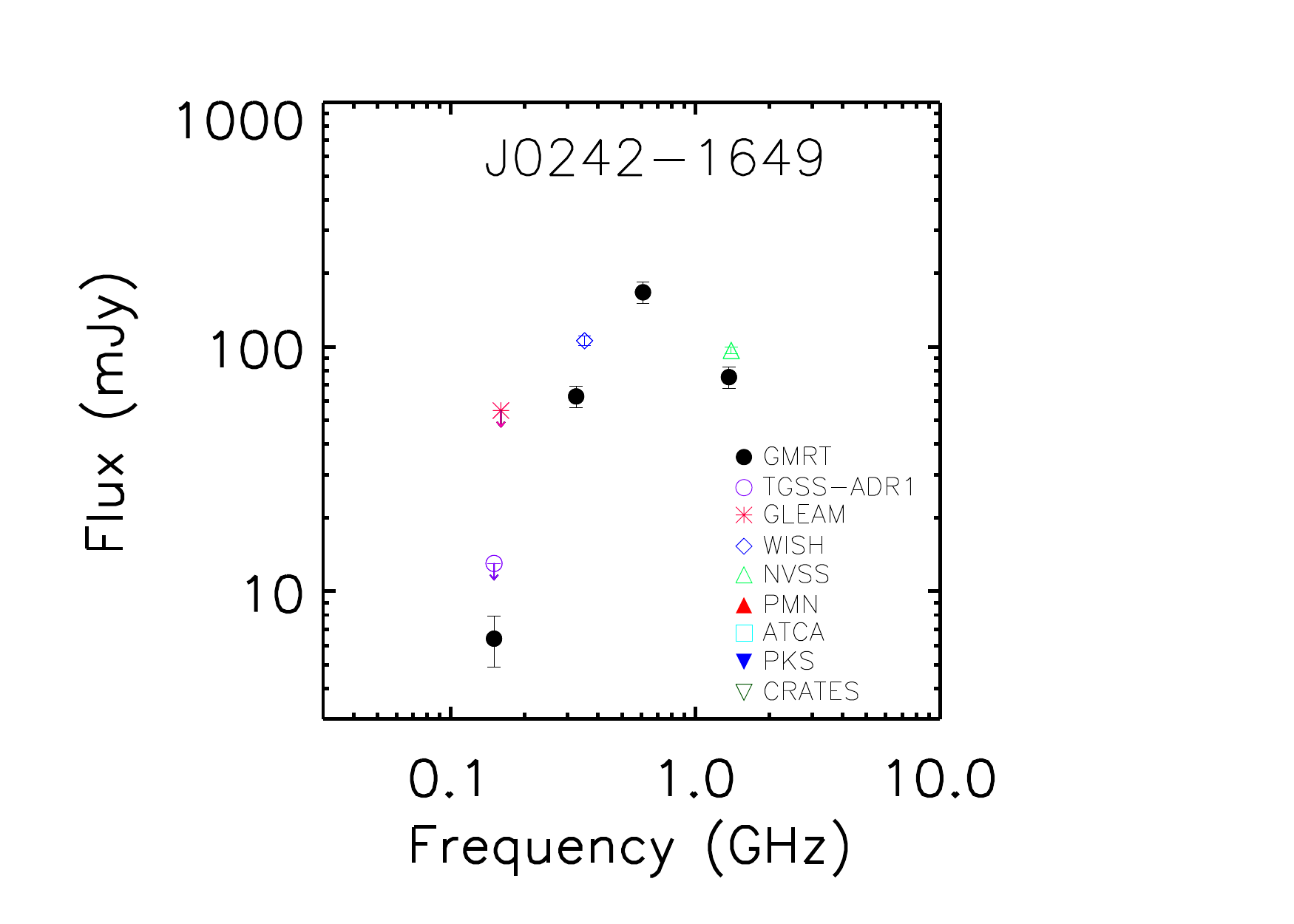}} 
\includegraphics[scale=1.1]{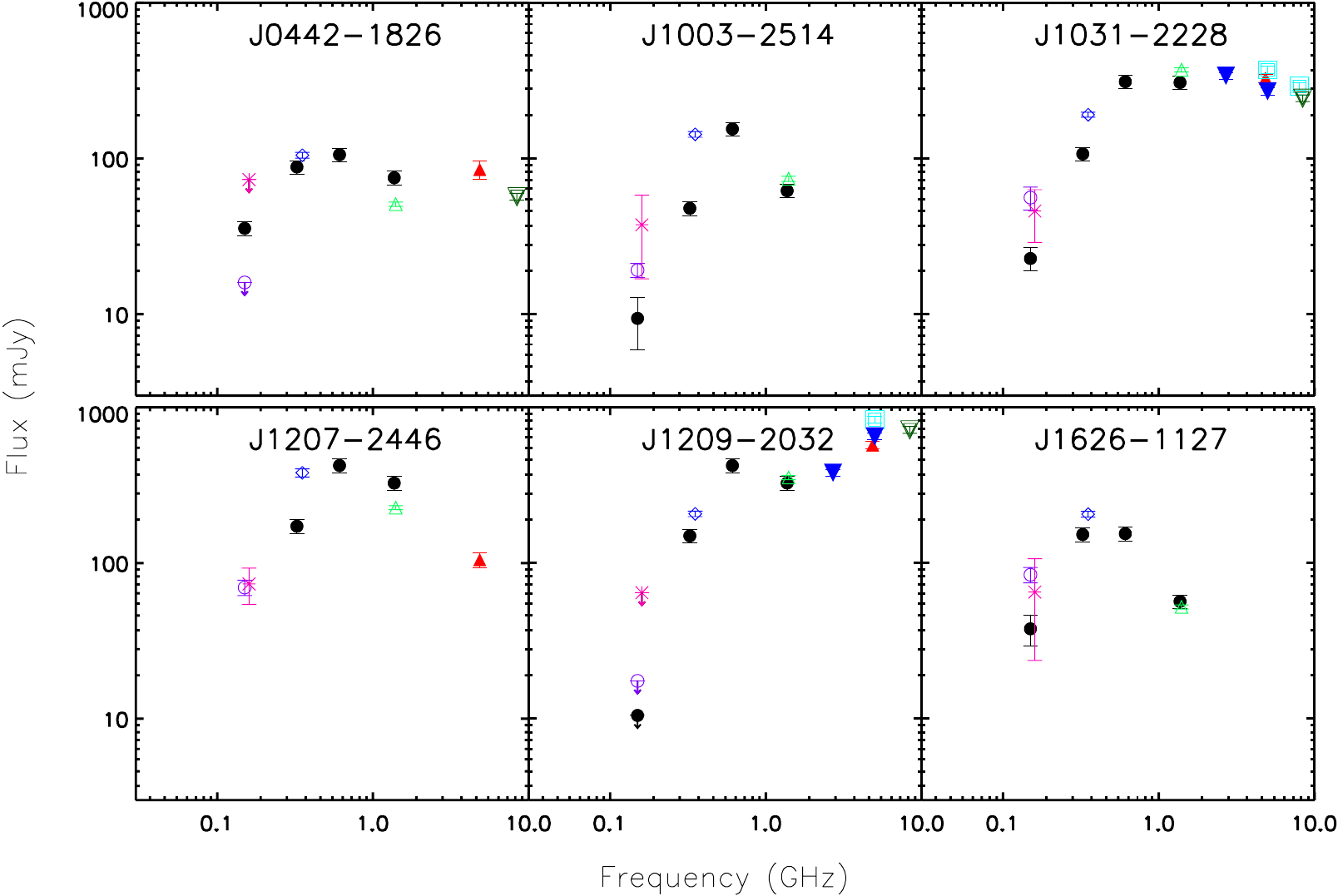} 
\caption{Radio spectra of the 7 target sources (see Table ~\ref{table:spec-prop}). In each panel, the names of the surveys are mentioned next to the corresponding symbol, with the frequency of the survey increasing downwards. The filled circles represent the measurements from the present GMRT observations at the 4 frequencies.} \label{fig:spec_all} 
\end{figure*}

\section{Discussion}
Based on the estimates of two-frequency spectral index $\alpha$(150-325 MHz) (Table~\ref{table:spec-prop}), we confirm J0242$-$1649 with $\alpha$ = $+$2.95$\pm$0.31 to be a bona-fide EISERS. 
Another source, J1209-2032 having $\alpha$ $>$ +2.64$\pm$0.13 is a promising EISERS candidate, i.e., $\alpha$ marginally above the SSA limit of $\alpha_{c}$ $=$ +$2.5$. None of the remaining 5 target 
sources is found to approach $\alpha_{c}$, although J1003$-$2514 and J1031$-$2228 do have very steeply inverted spectra, with $\alpha$ (150-325 MHz) $\sim$ +2.0. To verify if they are 
genuine EISERS, spectral measurements below 150 MHz will be helpful. It is encouraging that the two best cases of EISERS, emerging from the present work, are the same two that we had highlighted in 
Paper I.    

Recently, CEG17 have reported a few promising examples of EISERS based on the GLEAM survey which has catalogued nearly $10^{5}$ unresolved radio sources south of declination +30 deg and above a flux 
density limit of 0.16 Jy at 200 MHz, the effective frequency of the \lq deep wide-band images\rq~formed by integrating over the 170-231 MHz bandwidth \citep{Hurley-Walker2017,Wayth2015}. Note that 
the frequency of the present GMRT observations (and also of the TGSS-ADR1 \citep{Intema2017}) is almost a perfect match to the central frequency of the GLEAM survey (151 MHz). The key merit of the 
GLEAM survey is that it provides for each source essentially contemporaneous flux density measurements at several frequencies in the range 72 -- 231 MHz. The angular resolution, however, is modest 
($\sim$ 2 arcmin at 200 MHz) and, more importantly, the rms noise is usually much higher, starting with $\sim$ 20 mJy/beam at 231 MHz and rising to $\sim$ 100 mJy at 72 MHz. This latter point is 
specially pertinent for the discussion below, where we compare the findings of CEG17 with the present results.
  
Table 3 of CEG17 lists six radio sources for which their estimates of spectral index ($\alpha_{thick}$) in the optically thick region are close to, or exceed the SSA limit of $\alpha_{c}$ = +2.5. 
For two of them (J144815$-$162024 and J211949$-$343936), $\alpha_{thick}$ is actually consistent with the SSA limit of +2.5 (Table 3). For another source, J001513$-$472706, the 
GLEAM measurements at the lowest frequencies are too noisy to effectively constrain its spectral slope in the optically thick region. Likewise, difficulty arises for the source J100256$-$354157 
because of the poor fit of the theoretical spectrum to the flux measurements. For the remaining two sources, J074211$-$673309 and J213024$-$434819, CEG17 have estimated $\alpha_{thick}$ to be 
4.1$\pm$0.9 and 3.2$\pm$0.6, respectively. One aught to appreciate the model dependent nature of these estimates, since they actually represent the terminal slope of the theoretical synchrotron 
spectrum of a uniform synchrotron source (Eqn. 3 of CEG17), which CEG17 have fitted to their spectral measurements between 72 to 843/1400 MHz (note also that the points at 843 and 1400 MHz actually 
come from the observations made more than a decade prior to the GLEAM survey which only covers the 72 - 231 MHz band). 

In view of this, we have independently determined for the above two sources the slope of the inverted radio spectrum, by a weighted least-square fitting of a straight line to the GLEAM flux density 
measurements at their {\it lowest} (up to) 4 frequencies (since the effect of SSA is expected to be most pronounced at the lowest frequencies). These computed values, given in Table~\ref{table:GLEAM},
show that for J074211$-$673309, the uncertainty in the slope of the fitted spectrum is too large to infer a violation of the SSA limit of $\alpha_{thick}$ = +2.5. On the other hand, our best-fit 
spectral slope for J213024$-$434819 breaches the SSA limit by a significant margin, in accord with the claim of CEG17 (although they favour the FFA interpretation for the spectral turnover, on 
statistical ground). 

Here it may be cautioned that the present linear fitting to the GLEAM data points is not strictly valid and is merely intended to identify the most promising EISERS candidates. The problem 
relates to the correlated nature of the data in the GLEAM sub-bands \citep[see,][]{Hurley-Walker2017}, which means that the GLEAM measurements for individual sub-bands (such as the last four data 
points mentioned above) are not fully independent and, consequently, systematics can drive correlation and influence the measurements reproduced in Table~\ref{table:GLEAM} (potentially making a spectrum 
steeper or shallower). Bearing this in mind, if one assumes that the computed spectral slope of J213024$-$434819, which is based on the GLEAM measurement (Table~\ref{table:GLEAM}, will be confirmed by 
independent observations, both J0242$-$1649 (present work; Paper I) and J213024$-$434819 remain the best EISERS candidates known. A third good candidate is J1209$-$2032 ($\alpha$ $>$ 2.64$\pm$0.13 
between 150 and 325 MHz) which is identified with a galaxy at z$=$0.404 (Table~\ref{table:spec-prop}). It does not have a GPS type radio spectrum, unlike J0242$-$1649 (Figure~\ref{fig:spec_all}). 
As mentioned in sect. 4, its VLBI observations at 2.3 GHz show it to be a very young radio source, with a total extent of only 25 msec (i.e., $\sim$140 parsec at z$=$0.404). For such young radio 
sources, the observed sharp spectral turnover at metre wavelengths is also consistent with FFA occurring in a warm inhomogeneous interstellar medium plausibly associated with the host elliptical 
galaxy \citep{Bicknell2018}.

\begin{table}
\caption{The GLEAM survey flux densities and the estimated spectral indices$^{\star}$ for J074211$-$673309 and J213024$-$434819.}
\label{table:GLEAM}
\begin{tabular}{lcc}\\
\hline
\multicolumn{1}{l}{Parameters} & \multicolumn{1}{c}{J074211$-$673309} & \multicolumn{1}{c}{J213024$-$434819}\\
\hline
S$_{76}$ (mJy)$^{\dagger}$ 	&0.69$\pm$0.12 &0.15$\pm$0.07  	\\
S$_{84}$ (mJy)	&0.89$\pm$0.10 &0.27$\pm$0.05  	\\	
S$_{92}$ (mJy)	&1.02$\pm$0.09 &0.38$\pm$0.05  	\\
S$_{99}$ (mJy)	&1.14$\pm$0.10 &0.52$\pm$0.05  	\\
$\alpha_{(1)}$	&2.54$\pm$2.06 &5.70$\pm$4.79  	\\
$\alpha_{(2)}$	&1.89$\pm$0.34 &4.28$\pm$0.60	\\
$\alpha_{(3)}$	&1.70$\pm$0.19 &4.26$\pm$0.23	\\
\hline
\end{tabular}

{$\star$ The spectral indices represent slope of the straight line fitted to the flux densities (and errors) by the
weighted least-square method (see text).}\\
{$\dagger$ Flux densities at 76, 84, 92 and 99 MHz, taken from the GLEAM survey (CEG17).}\\
{$\alpha_{(1)}$ is the spectral index, based on the fit to the flux densities at the lowest two frequencies (sect. 5).}\\
{$\alpha_{(2)}$ is the spectral index, based on the fit to the flux densities at the lowest three frequencies (sect. 5).} \\
{$\alpha_{(3)}$ is the spectral index, based on the fit to the flux densities at the lowest four frequencies (sect. 5).} \\

\end{table}

\section{Conclusions}
This study presents our continued search for extragalactic sources whose radio continuum spectrum is so sharply inverted due to opacity effects that its slope $\alpha$ exceeds +2.5, thus becoming 
inconsistent with the standard interpretation in terms of synchrotron self-absorption (SSA) of the incoherent synchrotron radiation arising from relativistic particles with a power-law energy 
distribution. Here, we have reported sensitive, quasi-simultaneous GMRT observations at 150, 325, 610 and 1400 MHz of the seven sources we had previously identified as candidates for inverted radio 
spectrum which is steepened beyond the SSA limit (viz., EISERS). However, that surmise was based on published multi-frequency radio observations that are only moderately sensitive, as well as 
highly non-contemporaneous. Addressing both these shortcomings, our targeted GMRT observations at the four frequencies, as reported here, show that for at least one (probably, two) of the seven 
candidates, the spectral index computed between the lowest two frequencies, exceeds the SSA limit of $\alpha$ = +2.5. This, together with another one or two extragalactic sources that are reported 
in very recent literature to exhibit a similarly extreme radio spectrum (sect. 5; Table~\ref{table:GLEAM}), raises the possibility that in some very rare cases, the SSA limit may be violated.
However, if future observations indicate the presence of dense thermal plasma towards these extragalactic sources, that would render free-free absorption as a less radical and hence 
perhaps more attractive alternative explanation for the ultra-sharp turnover observed in their radio spectra. Search for additional sources with such extremely inverted radio continuum spectra forms 
a key objective of our ongoing programme, backed up with an ongoing follow-up using the upgraded GMRT \citep{Gupta2017}.
 
\section*{Acknowledgements}
The Giant Metrewave Radio Telescope (GMRT) is a national facility operated by the National Centre for Radio Astrophysics (NCRA) of the Tata Institute of Fundamental Research (TIFR).  We thank the 
staff at NCRA and GMRT for their support. This research has used the TIFR. GMRT. Sky. Survey (http://tgss.ncra.tifr.res.in) data products, NASA's Astrophysics Data System and NASA/IPAC Extragalactic 
Database (NED), Jet Propulsion Laboratory, California Institute of Technology under contract with National Aeronautics and Space Administration and VizieR catalogue access tool, CDS, Strasbourg, 
France. We thank the NRAO staff for providing \textsc{aips}. G-K thanks the National Academy of Sciences, India for the award of a NASI Senior Scientist Platinum Jubilee Fellowship. PD gratefully acknowledges generous support from the Indo-French Center for the Promotion of Advanced Research (Centre Franco-Indien pour la Promotion de la Recherche Avan\'{c}ee) under programme no. 5204-2 and IUCAA for its computational facilities.SP wants to thank DST INSPIRE Faculty Scheme (IF12/PH-44) for funding his research group. The authors would like to thank the anonymous referee for the valuable suggestions.

\bibliographystyle{mnras}
\bibliography{references}

\begin{thebibliography}{}
\makeatletter
\relax
\def\mn@urlcharsother{\let\do\@makeother \do\$\do\&\do\#\do\^\do\_\do\%\do\~}
\def\mn@doi{\begingroup\mn@urlcharsother \@ifnextchar [ {\mn@doi@}
  {\mn@doi@[]}}
\def\mn@doi@[#1]#2{\def\@tempa{#1}\ifx\@tempa\@empty \href
  {http://dx.doi.org/#2} {doi:#2}\else \href {http://dx.doi.org/#2} {#1}\fi
  \endgroup}
\def\mn@eprint#1#2{\mn@eprint@#1:#2::\@nil}
\def\mn@eprint@arXiv#1{\href {http://arxiv.org/abs/#1} {{\tt arXiv:#1}}}
\def\mn@eprint@dblp#1{\href {http://dblp.uni-trier.de/rec/bibtex/#1.xml}
  {dblp:#1}}
\def\mn@eprint@#1:#2:#3:#4\@nil{\def\@tempa {#1}\def\@tempb {#2}\def\@tempc
  {#3}\ifx \@tempc \@empty \let \@tempc \@tempb \let \@tempb \@tempa \fi \ifx
  \@tempb \@empty \def\@tempb {arXiv}\fi \@ifundefined
  {mn@eprint@\@tempb}{\@tempb:\@tempc}{\expandafter \expandafter \csname
  mn@eprint@\@tempb\endcsname \expandafter{\@tempc}}}

\bibitem[\protect\citeauthoryear{{An} \& {Baan}}{{An} \& {Baan}}{2012}]{An2012}
{An} T.,  {Baan} W.~A.,  2012, \mn@doi [\apj] {10.1088/0004-637X/760/1/77},
  \href {http://adsabs.harvard.edu/abs/2012ApJ...760...77A} {760, 77}

\bibitem[\protect\citeauthoryear{{Athreya} \& {Kapahi}}{{Athreya} \&
  {Kapahi}}{1998}]{Athreya1998}
{Athreya} R.~M.,  {Kapahi} V.~K.,  1998, \mn@doi [Journal of Astrophysics and
  Astronomy] {10.1007/BF02714911}, \href
  {http://adsabs.harvard.edu/abs/1998JApA...19...63A} {19, 63}

\bibitem[\protect\citeauthoryear{{Baars}, {Genzel}, {Pauliny-Toth}  \&
  {Witzel}}{{Baars} et~al.}{1977}]{Baars1977}
{Baars} J.~W.~M.,  {Genzel} R.,  {Pauliny-Toth} I.~I.~K.,   {Witzel} A.,  1977,
  \aap, \href {http://cdsads.u-strasbg.fr/abs/1977A%26A....61...99B} {61, 99}

\bibitem[\protect\citeauthoryear{{Bell} et~al.,}{{Bell}
  et~al.}{2019}]{Bell2019}
{Bell} M.~E.,  et~al., 2019, \mn@doi [\mnras] {10.1093/mnras/sty2801}, \href
  {http://adsabs.harvard.edu/abs/2019MNRAS.482.2484B} {482, 2484}

\bibitem[\protect\citeauthoryear{{Bicknell}, {Dopita}  \& {O'Dea}}{{Bicknell}
  et~al.}{1997}]{Bicknell1997}
{Bicknell} G.~V.,  {Dopita} M.~A.,   {O'Dea} C.~P.~O.,  1997, \mn@doi [\apj]
  {10.1086/304400}, \href {http://adsabs.harvard.edu/abs/1997ApJ...485..112B}
  {485, 112}

\bibitem[\protect\citeauthoryear{{Bicknell}, {Mukherjee}, {Wagner},
  {Sutherland}  \& {Nesvadba}}{{Bicknell} et~al.}{2018}]{Bicknell2018}
{Bicknell} G.~V.,  {Mukherjee} D.,  {Wagner} A.~Y.,  {Sutherland} R.~S.,
  {Nesvadba} N.~P.~H.,  2018, \mn@doi [\mnras] {10.1093/mnras/sty070}, \href
  {http://adsabs.harvard.edu/abs/2018MNRAS.475.3493B} {475, 3493}

\bibitem[\protect\citeauthoryear{{Callingham} et~al.,}{{Callingham}
  et~al.}{2015}]{Callingham2015}
{Callingham} J.~R.,  et~al., 2015, \mn@doi [\apj]
  {10.1088/0004-637X/809/2/168}, \href
  {http://adsabs.harvard.edu/abs/2015ApJ...809..168C} {809, 168}

\bibitem[\protect\citeauthoryear{{Callingham} et~al.,}{{Callingham}
  et~al.}{2017}]{Callingham2017}
{Callingham} J.~R.,  et~al., 2017, \mn@doi [\apj]
  {10.3847/1538-4357/836/2/174}, \href
  {http://adsabs.harvard.edu/abs/2017ApJ...836..174C} {836, 174}

\bibitem[\protect\citeauthoryear{{Chandra}, {Ray}  \& {Bhatnagar}}{{Chandra}
  et~al.}{2004}]{Chandra2004}
{Chandra} P.,  {Ray} A.,   {Bhatnagar} S.,  2004, \mn@doi [\apj]
  {10.1086/422675}, \href {http://adsabs.harvard.edu/abs/2004ApJ...612..974C}
  {612, 974}

\bibitem[\protect\citeauthoryear{{Condon}, {Cotton}, {Greisen}, {Yin},
  {Perley}, {Taylor}  \& {Broderick}}{{Condon} et~al.}{1998}]{Condon1998}
{Condon} J.~J.,  {Cotton} W.~D.,  {Greisen} E.~W.,  {Yin} Q.~F.,  {Perley}
  R.~A.,  {Taylor} G.~B.,   {Broderick} J.~J.,  1998, \mn@doi [\aj]
  {10.1086/300337}, \href {http://adsabs.harvard.edu/abs/1998AJ....115.1693C}
  {115, 1693}

\bibitem[\protect\citeauthoryear{{Coppejans}, {Cseh}, {Williams}, {van Velzen}
  \& {Falcke}}{{Coppejans} et~al.}{2015}]{Coppejans2015}
{Coppejans} R.,  {Cseh} D.,  {Williams} W.~L.,  {van Velzen} S.,   {Falcke} H.,
   2015, \mn@doi [\mnras] {10.1093/mnras/stv681}, \href
  {http://adsabs.harvard.edu/abs/2015MNRAS.450.1477C} {450, 1477}

\bibitem[\protect\citeauthoryear{{Coppejans} et~al.,}{{Coppejans}
  et~al.}{2016}]{Coppejans2016}
{Coppejans} R.,  et~al., 2016, \mn@doi [\mnras] {10.1093/mnras/stw799}, \href
  {http://adsabs.harvard.edu/abs/2016MNRAS.459.2455C} {459, 2455}

\bibitem[\protect\citeauthoryear{{Dallacasa}, {Stanghellini}, {Centonza}  \&
  {Fanti}}{{Dallacasa} et~al.}{2000}]{Dallacasa2000}
{Dallacasa} D.,  {Stanghellini} C.,  {Centonza} M.,   {Fanti} R.,  2000, \aap,
  \href {http://adsabs.harvard.edu/abs/2000A%26A...363..887D} {363, 887}

\bibitem[\protect\citeauthoryear{{De Breuck}, {Tang}, {de Bruyn},
  {R{\"o}ttgering}  \& {van Breugel}}{{De Breuck} et~al.}{2002}]{DeBreuck2002}
{De Breuck} C.,  {Tang} Y.,  {de Bruyn} A.~G.,  {R{\"o}ttgering} H.,   {van
  Breugel} W.,  2002, \mn@doi [\aap] {10.1051/0004-6361:20021115}, \href
  {http://adsabs.harvard.edu/abs/2002A%26A...394...59D} {394, 59}

\bibitem[\protect\citeauthoryear{{Falcke} et~al.,}{{Falcke}
  et~al.}{1999}]{Falcke1999}
{Falcke} H.,  et~al., 1999, \mn@doi [\apjl] {10.1086/311937}, \href
  {http://adsabs.harvard.edu/abs/1999ApJ...514L..17F} {514, L17}

\bibitem[\protect\citeauthoryear{{Falcke}, {K{\"o}rding}  \& {Nagar}}{{Falcke}
  et~al.}{2004}]{Falcke2004}
{Falcke} H.,  {K{\"o}rding} E.,   {Nagar} N.~M.,  2004, \mn@doi [\nar]
  {10.1016/j.newar.2004.09.029}, \href
  {http://adsabs.harvard.edu/abs/2004NewAR..48.1157F} {48, 1157}

\bibitem[\protect\citeauthoryear{{Gopal-Krishna} \&
  {Spoelstra}}{{Gopal-Krishna} \& {Spoelstra}}{1993}]{GopalKrishna1993}
{Gopal-Krishna} {Spoelstra} T.~A.~T.,  1993, \aap, \href
  {http://adsabs.harvard.edu/abs/1993A%26A...271..101G} {271, 101}

\bibitem[\protect\citeauthoryear{{Gopal-Krishna}, {Patnaik}  \&
  {Steppe}}{{Gopal-Krishna} et~al.}{1983}]{GopalKrishna1983}
{Gopal-Krishna} {Patnaik} A.~R.,   {Steppe} H.,  1983, \aap, \href
  {http://adsabs.harvard.edu/abs/1983A%26A...123..107G} {123, 107}

\bibitem[\protect\citeauthoryear{{Gopal-Krishna}, {Mhaskey}  \&
  {Mangalam}}{{Gopal-Krishna} et~al.}{2012}]{GopalKrishna2012}
{Gopal-Krishna} {Mhaskey} M.,   {Mangalam} A.,  2012, \mn@doi [\apj]
  {10.1088/0004-637X/744/1/31}, \href
  {http://adsabs.harvard.edu/abs/2012ApJ...744...31G} {744, 31}

\bibitem[\protect\citeauthoryear{{Gopal-Krishna}, {Sirothia}, {Mhaskey},
  {Ranadive}, {Wiita}, {Goyal}, {Kantharia}  \&
  {Ishwara-Chandra}}{{Gopal-Krishna} et~al.}{2014}]{GopalKrishna2014}
{Gopal-Krishna} {Sirothia} S.~K.,  {Mhaskey} M.,  {Ranadive} P.,  {Wiita}
  P.~J.,  {Goyal} A.,  {Kantharia} N.~G.,   {Ishwara-Chandra} C.~H.,  2014,
  \mn@doi [\mnras] {10.1093/mnras/stu1364}, \href
  {http://adsabs.harvard.edu/abs/2014MNRAS.443.2824K} {443, 2824}

\bibitem[\protect\citeauthoryear{{Griffith}, {Wright}, {Burke}  \&
  {Ekers}}{{Griffith} et~al.}{1994}]{Griffith1994}
{Griffith} M.~R.,  {Wright} A.~E.,  {Burke} B.~F.,   {Ekers} R.~D.,  1994,
  \mn@doi [\apjs] {10.1086/191863}, \href
  {http://adsabs.harvard.edu/abs/1994ApJS...90..179G} {90, 179}

\bibitem[\protect\citeauthoryear{Gupta et~al.,}{Gupta et~al.}{2017}]{Gupta2017}
Gupta Y.,  et~al., 2017, Current Science, 113, 707

\bibitem[\protect\citeauthoryear{{Hardcastle} et~al.,}{{Hardcastle}
  et~al.}{2016}]{Hardcastle2016}
{Hardcastle} M.~J.,  et~al., 2016, \mn@doi [\mnras] {10.1093/mnras/stw1763},
  \href {http://adsabs.harvard.edu/abs/2016MNRAS.462.1910H} {462, 1910}

\bibitem[\protect\citeauthoryear{{Healey} et~al.,}{{Healey}
  et~al.}{2008}]{Healey2008}
{Healey} S.~E.,  et~al., 2008, \mn@doi [\apjs] {10.1086/523302}, \href
  {http://adsabs.harvard.edu/abs/2008ApJS..175...97H} {175, 97}

\bibitem[\protect\citeauthoryear{{Hurley-Walker} et~al.,}{{Hurley-Walker}
  et~al.}{2017}]{Hurley-Walker2017}
{Hurley-Walker} N.,  et~al., 2017, \mn@doi [\mnras] {10.1093/mnras/stw2337},
  \href {http://adsabs.harvard.edu/abs/2017MNRAS.464.1146H} {464, 1146}

\bibitem[\protect\citeauthoryear{{Intema}}{{Intema}}{2014}]{Intema2014}
{Intema} H.~T.,  2014, in Astronomical Society of India Conference Series.
  (\mn@eprint {arXiv} {1402.4889})

\bibitem[\protect\citeauthoryear{{Intema}, {Jagannathan}, {Mooley}  \&
  {Frail}}{{Intema} et~al.}{2017}]{Intema2017}
{Intema} H.~T.,  {Jagannathan} P.,  {Mooley} K.~P.,   {Frail} D.~A.,  2017,
  \mn@doi [\aap] {10.1051/0004-6361/201628536}, \href
  {http://adsabs.harvard.edu/abs/2017A%26A...598A..78I} {598, A78}

\bibitem[\protect\citeauthoryear{{Jones} et~al.,}{{Jones}
  et~al.}{1996}]{Jones1996}
{Jones} D.~L.,  et~al., 1996, \mn@doi [\apjl] {10.1086/310183}, \href
  {http://adsabs.harvard.edu/abs/1996ApJ...466L..63J} {466, L63}

\bibitem[\protect\citeauthoryear{{Jones}, {Wehrle}, {Piner}  \&
  {Meier}}{{Jones} et~al.}{2001}]{Jones2001}
{Jones} D.~L.,  {Wehrle} A.~E.,  {Piner} B.~G.,   {Meier} D.~L.,  2001, \mn@doi
  [\apj] {10.1086/320979}, \href
  {http://adsabs.harvard.edu/abs/2001ApJ...553..968J} {553, 968}

\bibitem[\protect\citeauthoryear{{Kadler}, {Ros}, {Lobanov}, {Falcke}  \&
  {Zensus}}{{Kadler} et~al.}{2004}]{Kadler2004}
{Kadler} M.,  {Ros} E.,  {Lobanov} A.~P.,  {Falcke} H.,   {Zensus} J.~A.,
  2004, \mn@doi [\aap] {10.1051/0004-6361:20041051}, \href
  {http://adsabs.harvard.edu/abs/2004A%26A...426..481K} {426, 481}

\bibitem[\protect\citeauthoryear{{Kameno}, {Horiuchi}, {Shen}, {Inoue},
  {Kobayashi}, {Hirabayashi}  \& {Murata}}{{Kameno} et~al.}{2000}]{Kameno2000}
{Kameno} S.,  {Horiuchi} S.,  {Shen} Z.-Q.,  {Inoue} M.,  {Kobayashi} H.,
  {Hirabayashi} H.,   {Murata} Y.,  2000, \mn@doi [\pasj]
  {10.1093/pasj/52.1.209}, \href
  {http://adsabs.harvard.edu/abs/2000PASJ...52..209K} {52, 209}

\bibitem[\protect\citeauthoryear{{Kardashev}}{{Kardashev}}{1962}]{Kardashev1962}
{Kardashev} N.~S.,  1962, \sovast, \href
  {http://adsabs.harvard.edu/abs/1962SvA.....6..317K} {6, 317}

\bibitem[\protect\citeauthoryear{{Kellermann}}{{Kellermann}}{1964}]{Kellermann1964}
{Kellermann} K.~I.,  1964, \mn@doi [\apj] {10.1086/147998}, \href
  {http://adsabs.harvard.edu/abs/1964ApJ...140..969K} {140, 969}

\bibitem[\protect\citeauthoryear{{Kellermann}}{{Kellermann}}{1966}]{Kellermann1966}
{Kellermann} K.~I.,  1966, \mn@doi [Australian Journal of Physics]
  {10.1071/PH660195}, \href {http://adsabs.harvard.edu/abs/1966AuJPh..19..195K}
  {19, 195}

\bibitem[\protect\citeauthoryear{{Kellermann} \& {Pauliny-Toth}}{{Kellermann}
  \& {Pauliny-Toth}}{1969}]{Kellermann1969}
{Kellermann} K.~I.,  {Pauliny-Toth} I.~I.~K.,  1969, \mn@doi [\apjl]
  {10.1086/180305}, \href {http://adsabs.harvard.edu/abs/1969ApJ...155L..71K}
  {155, L71}

\bibitem[\protect\citeauthoryear{{Klamer}, {Ekers}, {Bryant}, {Hunstead},
  {Sadler}  \& {De Breuck}}{{Klamer} et~al.}{2006}]{Klamer2006}
{Klamer} I.~J.,  {Ekers} R.~D.,  {Bryant} J.~J.,  {Hunstead} R.~W.,  {Sadler}
  E.~M.,   {De Breuck} C.,  2006, \mn@doi [\mnras]
  {10.1111/j.1365-2966.2006.10714.x}, \href
  {http://adsabs.harvard.edu/abs/2006MNRAS.371..852K} {371, 852}

\bibitem[\protect\citeauthoryear{{Krichbaum}, {Alef}, {Witzel}, {Zensus},
  {Booth}, {Greve}  \& {Rogers}}{{Krichbaum} et~al.}{1998}]{Krichbaum1998}
{Krichbaum} T.~P.,  {Alef} W.,  {Witzel} A.,  {Zensus} J.~A.,  {Booth} R.~S.,
  {Greve} A.,   {Rogers} A.~E.~E.,  1998, \aap, \href
  {http://adsabs.harvard.edu/abs/1998A%26A...329..873K} {329, 873}

\bibitem[\protect\citeauthoryear{{Kuncic}, {Bicknell}  \& {Dopita}}{{Kuncic}
  et~al.}{1998}]{Kuncic1998}
{Kuncic} Z.,  {Bicknell} G.~V.,   {Dopita} M.~A.,  1998, \mn@doi [\apjl]
  {10.1086/311202}, \href {http://adsabs.harvard.edu/abs/1998ApJ...495L..35K}
  {495, L35}

\bibitem[\protect\citeauthoryear{{Levinson}, {Laor}  \& {Vermeulen}}{{Levinson}
  et~al.}{1995}]{Levinson1995}
{Levinson} A.,  {Laor} A.,   {Vermeulen} R.~C.,  1995, \mn@doi [\apj]
  {10.1086/175988}, \href {http://adsabs.harvard.edu/abs/1995ApJ...448..589L}
  {448, 589}

\bibitem[\protect\citeauthoryear{{Mangalam} \& {Gopal-Krishna}}{{Mangalam} \&
  {Gopal-Krishna}}{1995}]{Mangalam1995}
{Mangalam} A.~V.,  {Gopal-Krishna} 1995, \mn@doi [\mnras]
  {10.1093/mnras/275.4.976}, \href
  {http://adsabs.harvard.edu/abs/1995MNRAS.275..976M} {275, 976}

\bibitem[\protect\citeauthoryear{{Matveenko}, {Pauliny-Toth}  \&
  {Sherwood}}{{Matveenko} et~al.}{1990}]{Matveenko1990}
{Matveenko} L.~I.,  {Pauliny-Toth} I.~I.~K.,   {Sherwood} W.,  1990, Soviet
  Astronomy Letters, \href {http://adsabs.harvard.edu/abs/1990SvAL...16..247M}
  {16, 247}

\bibitem[\protect\citeauthoryear{{Murgia}, {Fanti}, {Fanti}, {Gregorini},
  {Klein}, {Mack}  \& {Vigotti}}{{Murgia} et~al.}{2002}]{Murgia2002}
{Murgia} M.,  {Fanti} C.,  {Fanti} R.,  {Gregorini} L.,  {Klein} U.,  {Mack}
  K.-H.,   {Vigotti} M.,  2002, \mn@doi [\nar] {10.1016/S1387-6473(01)00200-7},
  \href {http://adsabs.harvard.edu/abs/2002NewAR..46..307M} {46, 307}

\bibitem[\protect\citeauthoryear{{Murphy} et~al.,}{{Murphy}
  et~al.}{2010}]{Murphy2010}
{Murphy} T.,  et~al., 2010, \mn@doi [\mnras]
  {10.1111/j.1365-2966.2009.15961.x}, \href
  {http://adsabs.harvard.edu/abs/2010MNRAS.402.2403M} {402, 2403}

\bibitem[\protect\citeauthoryear{{O'Dea}}{{O'Dea}}{1998}]{Odea1998}
{O'Dea} C.~P.,  1998, \mn@doi [\pasp] {10.1086/316162}, \href
  {http://adsabs.harvard.edu/abs/1998PASP..110..493O} {110, 493}

\bibitem[\protect\citeauthoryear{{Ostorero} et~al.,}{{Ostorero}
  et~al.}{2010}]{Ostorero2010}
{Ostorero} L.,  et~al., 2010, \mn@doi [\apj] {10.1088/0004-637X/715/2/1071},
  \href {http://adsabs.harvard.edu/abs/2010ApJ...715.1071O} {715, 1071}

\bibitem[\protect\citeauthoryear{{Pacholczyk}}{{Pacholczyk}}{1970}]{Pacholczyk1970}
{Pacholczyk} A.~G.,  1970, {Radio astrophysics. Nonthermal processes in
  galactic and extragalactic sources}

\bibitem[\protect\citeauthoryear{{Perley} \& {Butler}}{{Perley} \&
  {Butler}}{2017}]{Perley2017}
{Perley} R.~A.,  {Butler} B.~J.,  2017, \mn@doi [\apjs]
  {10.3847/1538-4365/aa6df9}, \href
  {http://adsabs.harvard.edu/abs/2017ApJS..230....7P} {230, 7}

\bibitem[\protect\citeauthoryear{{Rees}}{{Rees}}{1967}]{Rees1967}
{Rees} M.~J.,  1967, \mn@doi [\mnras] {10.1093/mnras/136.3.279}, \href
  {http://adsabs.harvard.edu/abs/1967MNRAS.136..279R} {136, 279}

\bibitem[\protect\citeauthoryear{{Rengelink}, {Tang}, {de Bruyn}, {Miley},
  {Bremer}, {Roettgering}  \& {Bremer}}{{Rengelink}
  et~al.}{1997}]{Rengelink1997}
{Rengelink} R.~B.,  {Tang} Y.,  {de Bruyn} A.~G.,  {Miley} G.~K.,  {Bremer}
  M.~N.,  {Roettgering} H.~J.~A.,   {Bremer} M.~A.~R.,  1997, \mn@doi [\aaps]
  {10.1051/aas:1997358}, \href
  {http://adsabs.harvard.edu/abs/1997A%26AS..124..259R} {124, 259}

\bibitem[\protect\citeauthoryear{{Roger}, {Costain}  \& {Bridle}}{{Roger}
  et~al.}{1973}]{Roger1973}
{Roger} R.~S.,  {Costain} C.~H.,   {Bridle} A.~H.,  1973, \mn@doi [\aj]
  {10.1086/111506}, \href
  {https://ui.adsabs.harvard.edu/\#abs/1973AJ.....78.1030R} {78, 1030}

\bibitem[\protect\citeauthoryear{{Rybicki} \& {Lightman}}{{Rybicki} \&
  {Lightman}}{1986}]{Rybicki1979}
{Rybicki} G.~B.,  {Lightman} A.~P.,  1986, {Radiative Processes in
  Astrophysics}

\bibitem[\protect\citeauthoryear{{Scheuer} \& {Williams}}{{Scheuer} \&
  {Williams}}{1968}]{Scheuer1968}
{Scheuer} P.~A.~G.,  {Williams} P.~J.~S.,  1968, \mn@doi [\araa]
  {10.1146/annurev.aa.06.090168.001541}, \href
  {http://adsabs.harvard.edu/abs/1968ARA%26A...6..321S} {6, 321}

\bibitem[\protect\citeauthoryear{{Slish}}{{Slish}}{1963}]{Slish1963}
{Slish} V.~I.,  1963, \mn@doi [\nat] {10.1038/199682a0}, \href
  {http://adsabs.harvard.edu/abs/1963Natur.199..682S} {199, 682}

\bibitem[\protect\citeauthoryear{{Spoelstra}, {Patnaik}  \&
  {Gopal-Krishna}}{{Spoelstra} et~al.}{1985}]{Spoelstra1985}
{Spoelstra} T.~A.~T.,  {Patnaik} A.~R.,   {Gopal-Krishna} 1985, \aap, \href
  {http://adsabs.harvard.edu/abs/1985A%26A...152...38S} {152, 38}

\bibitem[\protect\citeauthoryear{{Stawarz}, {Ostorero}, {Begelman}, {Moderski},
  {Kataoka}  \& {Wagner}}{{Stawarz} et~al.}{2008}]{Stawarz2008}
{Stawarz} {\L}.,  {Ostorero} L.,  {Begelman} M.~C.,  {Moderski} R.,  {Kataoka}
  J.,   {Wagner} S.,  2008, \mn@doi [\apj] {10.1086/587781}, \href
  {http://adsabs.harvard.edu/abs/2008ApJ...680..911S} {680, 911}

\bibitem[\protect\citeauthoryear{{Swarup}}{{Swarup}}{1991}]{Swarup1991}
{Swarup} G.,  1991, in {Cornwell} T.~J.,  {Perley} R.~A.,  eds,  Astronomical
  Society of the Pacific Conference Series Vol. 19, IAU Colloq. 131: Radio
  Interferometry. Theory, Techniques, and Applications. pp 376--380

\bibitem[\protect\citeauthoryear{{Taylor}, {Stil}  \& {Sunstrum}}{{Taylor}
  et~al.}{2009}]{Taylor2009}
{Taylor} A.~R.,  {Stil} J.~M.,   {Sunstrum} C.,  2009, \mn@doi [\apj]
  {10.1088/0004-637X/702/2/1230}, \href
  {http://adsabs.harvard.edu/abs/2009ApJ...702.1230T} {702, 1230}

\bibitem[\protect\citeauthoryear{{Tingay} \& {Murphy}}{{Tingay} \&
  {Murphy}}{2001}]{Tingay2001}
{Tingay} S.~J.,  {Murphy} D.~W.,  2001, \mn@doi [\apj] {10.1086/318247}, \href
  {http://adsabs.harvard.edu/abs/2001ApJ...546..210T} {546, 210}

\bibitem[\protect\citeauthoryear{{Tingay} \& {de Kool}}{{Tingay} \& {de
  Kool}}{2003}]{Tingay2003}
{Tingay} S.~J.,  {de Kool} M.,  2003, \mn@doi [\aj] {10.1086/376600}, \href
  {http://adsabs.harvard.edu/abs/2003AJ....126..723T} {126, 723}

\bibitem[\protect\citeauthoryear{{Vermeulen}, {Ros}, {Kellermann}, {Cohen},
  {Zensus}  \& {van Langevelde}}{{Vermeulen} et~al.}{2003}]{Vermeulen2003}
{Vermeulen} R.~C.,  {Ros} E.,  {Kellermann} K.~I.,  {Cohen} M.~H.,  {Zensus}
  J.~A.,   {van Langevelde} H.~J.,  2003, \mn@doi [\aap]
  {10.1051/0004-6361:20021752}, \href
  {http://adsabs.harvard.edu/abs/2003A%26A...401..113V} {401, 113}

\bibitem[\protect\citeauthoryear{{Walker}, {Dhawan}, {Romney}, {Kellermann}  \&
  {Vermeulen}}{{Walker} et~al.}{2000}]{Walker2000}
{Walker} R.~C.,  {Dhawan} V.,  {Romney} J.~D.,  {Kellermann} K.~I.,
  {Vermeulen} R.~C.,  2000, \mn@doi [\apj] {10.1086/308372}, \href
  {http://adsabs.harvard.edu/abs/2000ApJ...530..233W} {530, 233}

\bibitem[\protect\citeauthoryear{{Wayth} et~al.,}{{Wayth}
  et~al.}{2015}]{Wayth2015}
{Wayth} R.~B.,  et~al., 2015, \mn@doi [\pasa] {10.1017/pasa.2015.26}, \href
  {http://adsabs.harvard.edu/abs/2015PASA...32...25W} {32, e025}

\bibitem[\protect\citeauthoryear{{de Kool} \& {Begelman}}{{de Kool} \&
  {Begelman}}{1989}]{deKool1989}
{de Kool} M.,  {Begelman} M.~C.,  1989, \mn@doi [\nat] {10.1038/338484a0},
  \href {http://adsabs.harvard.edu/abs/1989Natur.338..484D} {338, 484}

\bibitem[\protect\citeauthoryear{{van Breugel}}{{van
  Breugel}}{1984}]{vanBreugel1984}
{van Breugel} W.,  1984, in {Fanti} R.,  {Kellermann} K.~I.,   {Setti} G.,
  eds,  IAU Symposium Vol. 110, VLBI and Compact Radio Sources. p.~59

\makeatother
\end{thebibliography}
\clearpage

\bsp	
\label{lastpage}
\end{document}


\begin{strip}
\thispagestyle{empty}
\begin{center}	
\vspace*{-3.0cm} 
{\textbf{ONLINE MATERIAL}}\\
{GMRT observations of extragalactic radio sources with steeply inverted spectra}\\
\vspace*{0.6cm} \small{{Mukul Mhaskey$^{1}$, Gopal-Krishna$^{2}$, Pratik Dabhade$^{3,4}$, Surajit Paul$^{1}$}\\
 {Sameer Salunkhe$^{1}$, and S.K. Sirothia$^{5,6}$}}\\
 
\vspace*{0.3cm} \it \tiny{$^{1}$Department of Physics, Savitribai Phule Pune Unversity, Ganeshkhind, Pune 411007, India\\
 $^{2}$Aryabhatta Research Institute of Observational Sciences (ARIES), Manora Peak, Nainital $-$ 263129, India\\
 $^{3}$Inter University Centre for Astronomy and Astrophysics (IUCAA), Pune 411007, India\\
 $^{4}$Leiden Observatory, Leiden University, Niels Bohrweg 2, 2333 CA, Leiden, Netherlands\\
 $^{5}$Square Kilometre Array South Africa, 3rd Floor, The Park, Park Road, Pinelands, 7405, South Africa\\
 $^{6}$Department of Physics and Electronics, Rhodes University, PO Box 94, Grahamstown, 6140, South Africa} \\
\end{center}
\end{strip}

\begin{figure}
\includegraphics[width=3.0in,height=3.0in]{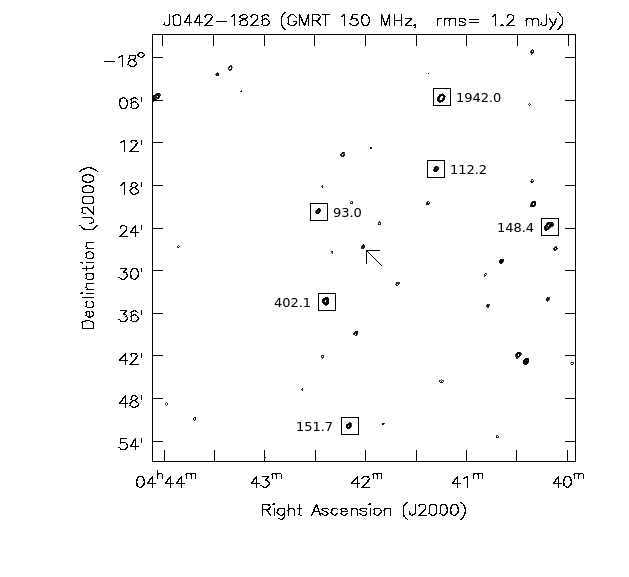} 
\includegraphics[width=3.0in,height=3.0in]{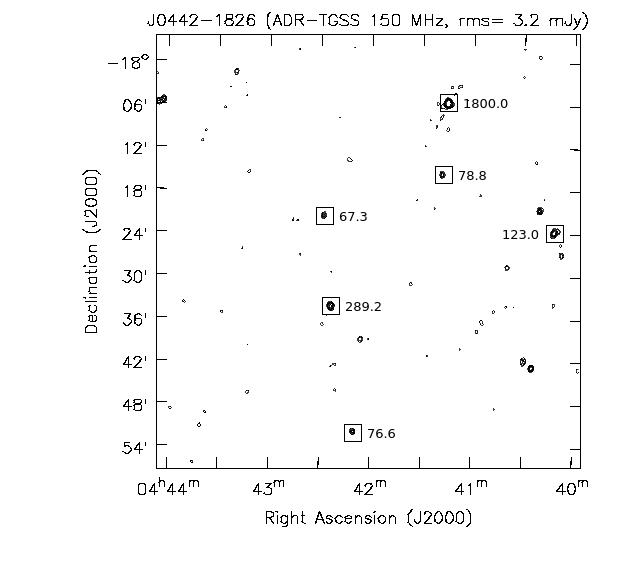} 
\caption{\scriptsize{Contour maps of J0442-1649 at 150 MHz from the present GMRT observations (upper panel) and the TGSS-ADR1 (lower panel). The contours in both the maps are plotted in the intensity units (mJy/beam) for comparison sake. The relative contour levels are 0.005, 0.010, 0.0.020, 0.040, 0.080, 0.160, 0.320 with the unit contour set to 1. The contours start at 5 mJy and double every level. The target source J0442-1649 ($\sim$ 35.6 mJy, Table 2) is seen at the centre of the present GMRT map (marked with arrow); however it is undetected in the ADR-TGSS map. The beam size is $27.2~\times~17.4{"}$ (PA= $-34^{\circ}$) at 150 MHz (GMRT) and 31.83 $\times~25.0^{"}$ at 150 MHz (TGSS-ADR1). The target source J0442-1649 ($\sim$ 35.6 mJy, Table 2) is seen at the centre of the present GMRT map; however it is undetected in the ADR-TGSS map. Flux densities (mJy), marked for several unresolved sources in the two maps, were used for calculating the flux scaling factor (FSF) between these maps (Table 3).}} \label{fig:J0442-with-labels}  
\end{figure}

\begin{figure}
\includegraphics[width=3.0in]{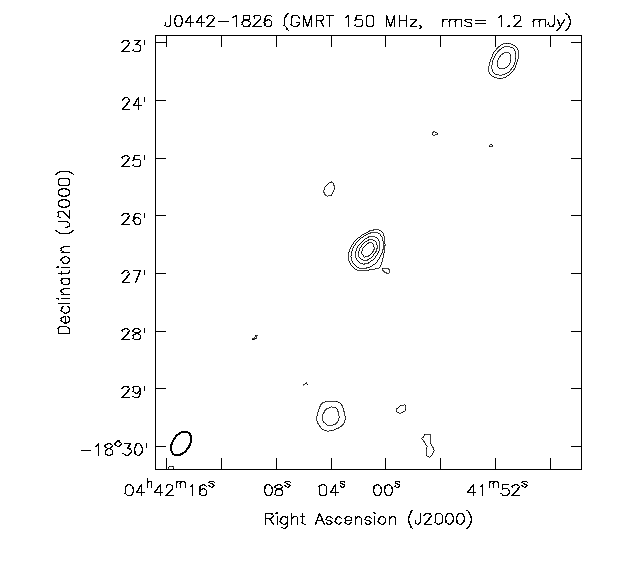} 
\includegraphics[width=3.0in]{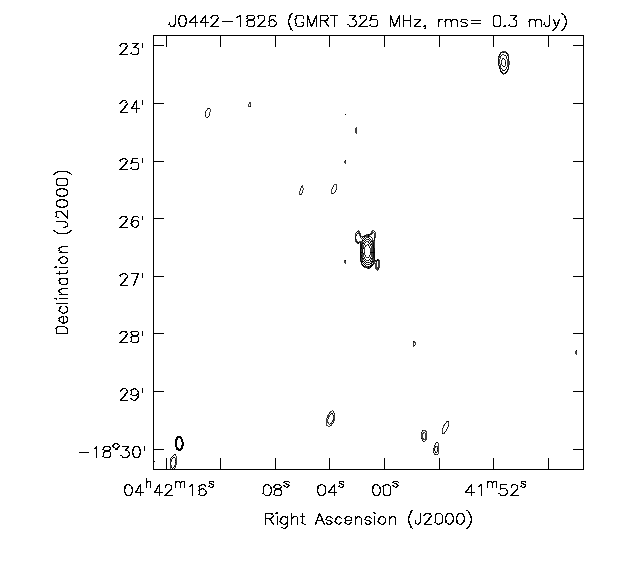} 
\caption{\scriptsize{GMRT contour maps of J0442-1826 at 150 MHz and 325 MHz, respectively. The contour levels are 3,4,8,16,32,64 \& 128 with the unit contour level at 150 MHz at 1.3 mJy and 0.3 mJy at 325 MHz. The beam size is $27.2~\times~17.4^{"}$ (PA= $-34^{\circ}$) and $13.7~\times~6.9{"}$ (PA= $2^{\circ}$) at 150 MHz and 325 MHz respectively. The target source lies at the centre of the map.}} \label{fig:J0442}  
\end{figure}

\begin{figure}
\includegraphics[width=3.0in]{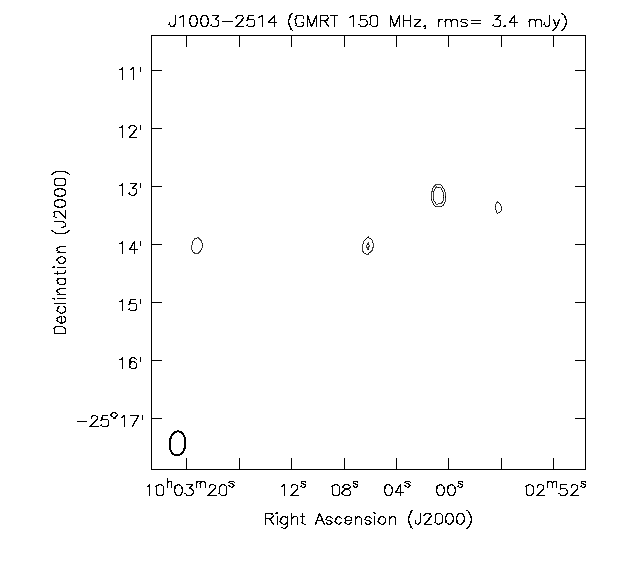} 
\includegraphics[width=3.0in]{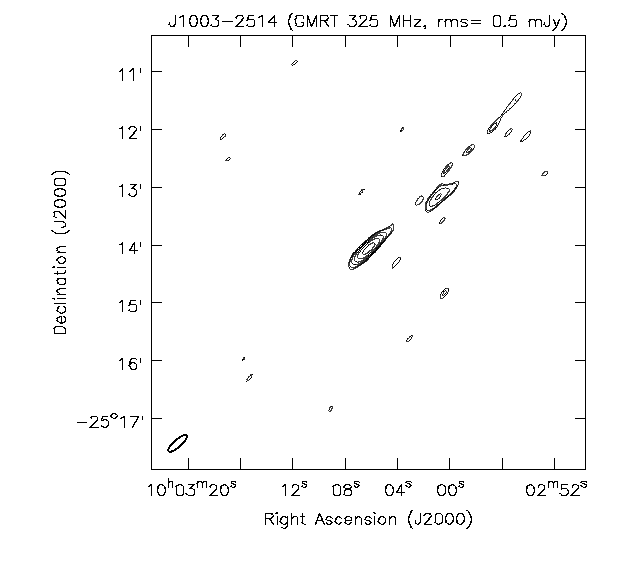} 
\caption{\scriptsize{GMRT contour maps of J1003-2514 at 150 MHz and 325 MHz, respectively. The contour levels are 3,4,8,16,32,64 \& 128 with the unit contour level at 150 MHz at 3.4 mJy and 0.5 mJy at 325 MHz. The beam size is $24.9~\times~16.4^{"}$ (PA= $-4^{\circ}$) and $24.6~\times~7.6{"}$ (PA= $-48^{\circ}$) at 150 MHz and 325 MHz respectively. The target source lies at the centre of the map.}} \label{fig:J1003} 
\end{figure}

\begin{figure}
\includegraphics[width=3.0in]{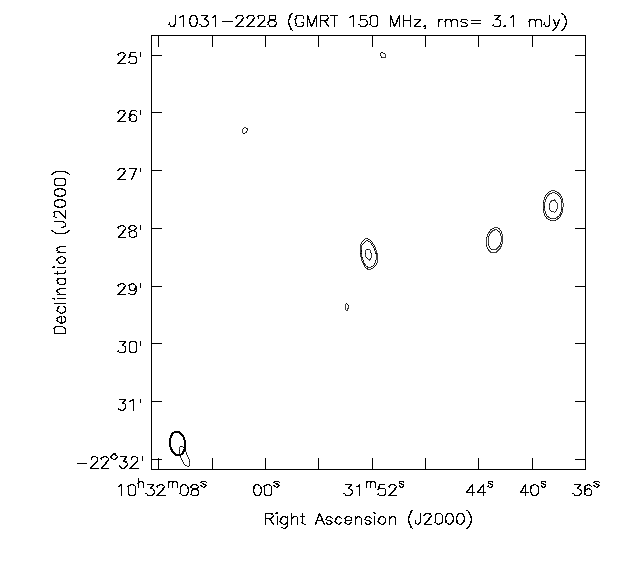} 
\includegraphics[width=3.0in]{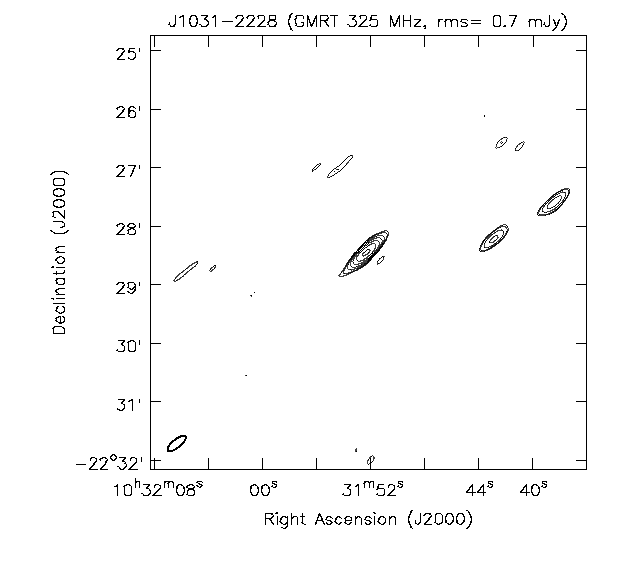} 
\caption{\scriptsize{GMRT contour maps of J1031-2228 at 150 MHz and 325 MHz, respectively. The contour levels are 3,4,8,16,32,64 \& 128 with the unit contour level at 150 MHz at 3.1 mJy and 0.7 mJy at 325 MHz. The beam size is $24.6~\times~15.7^{"}$ (PA= $7{\circ}$) and $23.1~\times~8.2{"}$ (PA= $-51^{\circ}$) at 150 MHz and 325 MHz respectively. The target source lies at the centre of the map.}} \label{fig:J1031}  
\end{figure}

\begin{figure}
\includegraphics[width=3.0in]{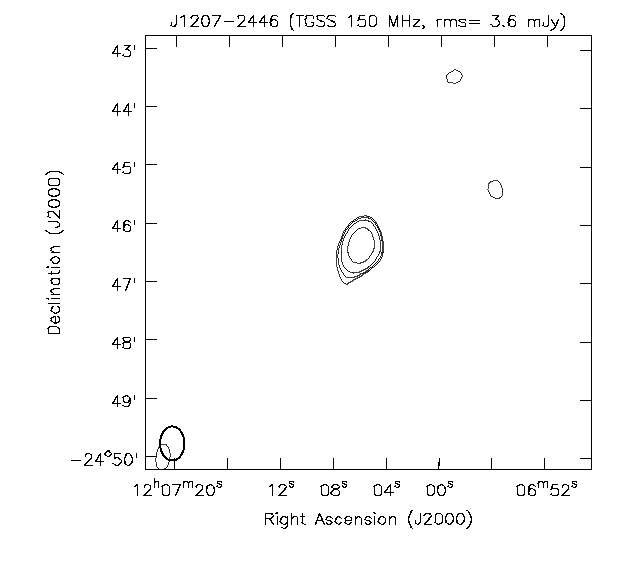} 
\includegraphics[width=3.0in]{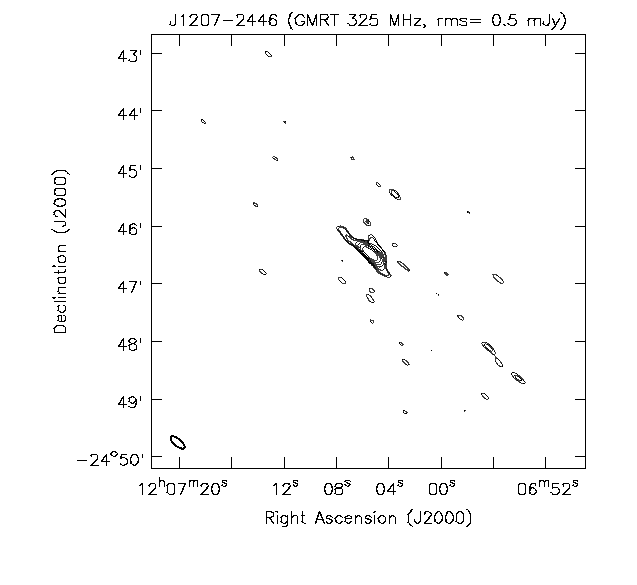} 
\caption{\scriptsize{GMRT contour maps of J1207-2446 at 150 MHz and 325 MHz, respectively. The contour levels are 3,4,8,16,32,64 \& 128 with the unit contour level at 150 MHz at 3.6 mJy and 0.5 mJy at 325 MHz. The beam size is $33.7~\times~12.8^{"}$ (PA= $42^{\circ}$) and $18.1~\times~7.3{"}$ (PA= $50^{\circ}$) at 150 MHz and 325 MHz respectively. The target source lies at the centre of the map}} \label{fig:J1207} 
\end{figure}

\begin{figure}
\includegraphics[width=3.0in]{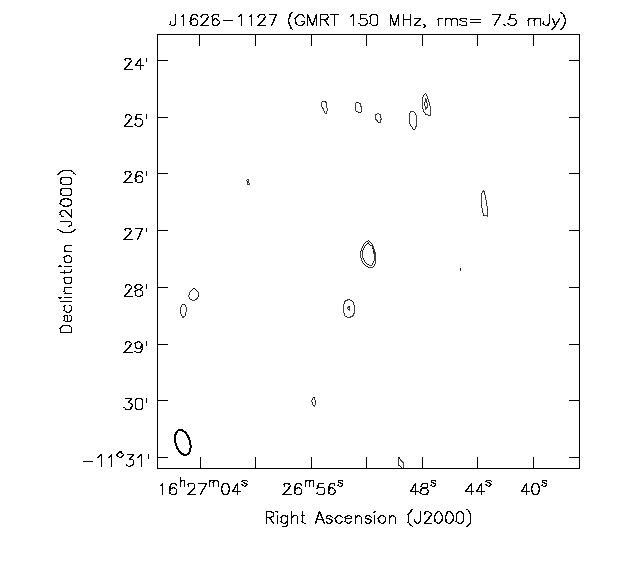} 
\includegraphics[width=3.0in]{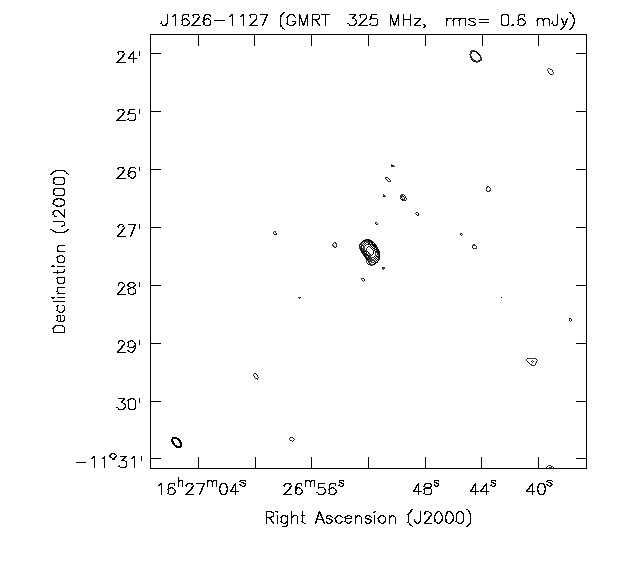} 
\caption{\scriptsize{GMRT contour maps of J1626-1127 at 150 MHz and 325 MHz, respectively. The contour levels are 3,4,8,16,32,64 \& 128 with the unit contour level at 150 MHz at 7.5 mJy and 0.6 mJy at 325 MHz. The beam size is $27.3~\times~15.1^{"}$ (PA= $17^{\circ}$) and $11.7~\times~6.8{"}$ (PA= $38^{\circ}$) at 150 MHz and 325 MHz respectively. The target source lies at the centre of the map.}} \label{fig:J1626} 
\end{figure}

\clearpage